\documentclass[aps,prd,onecolumn,groupedaddress,showpacs]{revtex4}

\usepackage{graphicx}
\usepackage{dcolumn}
\usepackage{bm}
\usepackage{amssymb}
\usepackage{amsmath}
\usepackage{epsfig}    
\usepackage{color}

\def\be{\begin{equation}}
\def\ee{\end{equation}}
\newcommand{\bea}{\begin{eqnarray}}
\newcommand{\eea}{\end{eqnarray}}

\newcommand{\nuh}{{\nu_{\rm H}}}

\def \HII {H{\sc \,ii}}

\numberwithin{equation}{section}

\begin{document} 
\title{Effects of Long-lived 10 MeV Scale Sterile Neutrino on Primordial Elemental Abundances and Effective Neutrino Number}
\preprint{KIAS-P14017, TU-963}

\author{Hiroyuki Ishida}
\email{h_ishida@tuhep.phys.tohoku.ac.jp}
\affiliation{Department of Physics, Tohoku University, Sendai 980-8578, Japan}
\author{Motohiko Kusakabe}
\email{motohiko@kau.ac.kr}
\affiliation{School of Liberal Arts and Science, Korea Aerospace University, Goyang 412-791, Korea}
\affiliation{Department of Physics, Soongsil University, Seoul 156-743, Korea}
\author{Hiroshi Okada}
\email{ hokada@kias.re.kr}
\affiliation{School of Physics, KIAS, Seoul 130-722, Korea}
%



\begin{abstract}
The primordial lithium abundance inferred from spectroscopic observations of metal-poor stars is $\sim 3$ times smaller than the theoretical prediction in standard big bang nucleosynthesis (BBN) model.  We assume a simple model composed of standard model particles and a sterile neutrino $\nuh$ with mass of $\mathcal{O}(10)$ MeV which decays long after BBN.  We then investigate cosmological effects of a sterile neutrino decay, and check if a sterile neutrino can reduce the primordial lithium abundance.
We formulate the injection spectrum of nonthermal photon induced by the $\nuh$ decay.  We take into account the generation of electrons and positrons, $e^\pm$'s, and active neutrinos at the $\nuh$ decay, the primary photon production via the inverse Compton scattering of cosmic background radiation (CBR) by energetic $e^\pm$, and electromagnetic cascade showers induced by the primary photons.  The steady state injection spectrum is then derived as a function of the $\nuh$ mass and the photon temperature.  
The $\nuh$ decay produces energetic active neutrinos which are not thermalized, and $e^\pm$'s which are thermalized.  We then derive formulae relevant to the $\nuh$ decay rates and formulae for the baryon-to-photon ratio $\eta$ and effective neutrino number $N_{\rm eff}$.  
The initial abundance, mass, and lifetime of $\nuh$ are taken as free parameters.  We then consistently solve (1) the cosmic thermal history, (2) nonthermal nucleosynthesis induced by the nonthermal photons, (3) the $\eta$ value, and (4) the $N_{\rm eff}$ value.  
We find that an effective $^7$Be destruction can occur only if the sterile neutrino decays at photon temperature $T={\cal O}(1)$ keV.  
Amounts of energy injection at the $\nuh$ decay are constrained from limits on primordial D and $^7$Li abundances, the $N_{\rm eff}$ value, and the CBR energy spectrum.  We find that $^7$Be is photodisintegrated and the Li problem is partially solved for the lifetime $10^4-10^5$ s and the mass $\gtrsim 14$ MeV.  $^7$Be destruction by more than a factor of three is not possible because of an associated D over-destruction.  In the parameter region, the $\eta$ value is decreased slightly, while the $N_{\rm eff}$ value is increased by a factor of $\Delta N_{\rm eff} \lesssim$ 1.
In this study, errors in photodisintegration cross sections of $^7$Be($\gamma$, $\alpha$)$^3$He and $^7$Li($\gamma$, $\alpha$)$^3$H that have propagated through literatures are corrected, and new functions are derived based on recent nuclear experiments.  It is found that the new photodisintegration rates are 2.3 to 2.5 times smaller than the old rates.  The correct cross sections thus indicate significantly smaller efficiencies of $^7$Be and $^7$Li photodisintegration.
Abundances of sterile neutrino necessary for the $^7$Li reduction are much smaller than thermal freeze-out abundances.  The relic sterile neutrino, therefore, must be diluted between the freeze-out and BBN epochs by some mechanism.
\end{abstract}


\maketitle
\section{Introduction\label{sec:intro}}
Big bang nucleosynthesis (BBN) model \cite{Alpher1948} successfully explains 
primordial light element abundances inferred from astronomical observations 
(e.g. \cite{Coc:2011az,Coc:2013eea}) if the cosmological baryon density determined from 
the power spectrum of cosmic microwave background (CMB) radiation measured with 
the Wilkinson Microwave Anisotropy Probe (WMAP) \cite{Spergel:2003cb,Spergel:2006hy,
Larson:2010gs,Hinshaw:2012fq} or Planck \cite{Ade:2013zuv} is adopted.  
An apparent discrepancy, however, exists between observational and theoretical 
$^7$Li abundances.  Spectroscopic observations of metal-poor stars (MPSs) 
indicate a plateau for the abundance ratio, $^7$Li/H$=(1-2) \times 10^{-10}$, 
with small error bars as a function of 
metallicity for [Fe/H]$>-3$ \footnote{[A/B]$=\log(n_{\rm A}/n_{\rm B})-\log(n_{\rm A}/n_{\rm B})_\odot$, where $n_i$ is the number density of element ($i$=A and B), and the subscript $\odot$ indicates the solar value.} in the Galaxy~\cite{Spite:1982dd,Ryan:1999vr,Melendez:2004ni,Asplund:2005yt,
bon2007,shi2007,Aoki:2009ce,Hernandez:2009gn,Sbordone:2010zi,Monaco:2011sd,Mucciarelli:2011ts,Aoki:2012wb,Aoki2012b} and $\omega$ Centauri accreted by the Galaxy~\cite{Monaco:2010mm} \footnote{Average stellar Li abundances in metal-poor globular clusters (GCs; e.g. \cite{Hernandez:2009gn,Mucciarelli:2010gz}) are larger than those in metal-poor halo stars.   The Li abundance in GC M4 turn-off stars has been determined to be log($^7$Li/H)$=-12+(2.30\pm 0.02 \pm 0.10)$ \cite{Mucciarelli:2010gz}, while the Li abundance in halo dwarf stars is log($^7$Li/H)$=-12+(2.199\pm 0.086)$ \cite{Sbordone:2010zi}.  However, these abundances are consistent with each other within the uncertainties for the moment (see fig. 3 of Ref. \cite{Mucciarelli:2010gz}).  The possible systematic difference in the Li abundances in GCs and Galactic halo should be studied further in future.}.  The plateau abundance is $\sim 3-4$ times smaller than that predicted in standard BBN (SBBN) model (e.g., $^7$Li/H=$5.24 \times 10^{-10}$~\cite{Coc:2011az}; 
see Ref. \cite{Coc:2013eea} for theoretical light element abundances for the baryon 
density from Planck \cite{Ade:2013zuv}).  Recent observations also indicate a break of this plateau shown as small Li abundances with large dispersion at lower metallicities of [Fe/H]$<-3$ \cite{Frebel:2005ig,Aoki:2005gn,bon2007,Aoki:2009ce,Hernandez:2009gn,Sbordone:2010zi,Aoki2012b}.  Therefore, it seems that we need a mechanism for a metallicity-independent depletion from the primordial abundance to the plateau abundance and also another for a metallicity-dependent depletion from the plateau abundance.  In this paper, we focus on the former universal depletion by cosmological processes.
This Li problem (see Ref. \cite{Fields:2011zzb} for a review) shows that some physical 
processes have reduced the primordial Li abundance in some epoch during or after BBN.

Standard stellar model suggests very small depletions of Li isotopes 
in surfaces of MPSs \cite{Deliyannis1990}. The $^7$Li/H abundances of MPSs observed 
today are then approximately interstellar abundances when the MPSs formed. 
Nonstandard processes such as the rotationally induced mixing
\cite{Pinsonneault:1998nf,Pinsonneault:2001ub}, and the turbulent
mixing~\cite{Richard:2004pj,Korn:2007cx,Lind:2009ta} have been suggested to reduce
the $^7$Li abundance in stellar atmospheres. 
In the former model, a large depletion factor does not realize simultaneously with a small dispersion in stellar Li abundances after the depletion.  The depletion factor is then constrained to be small.
In the latter model, a depletion
of a factor of $1.6-2.0$ \cite{Richard:2004pj} is predicted although it is still unclear
if this mechanism can deplete Li abundances of all MPSs rather uniformly.

Nonstandard BBN, on the other hand, may be responsible for the Li problem at least partially.
We note that $^7$Be is produced more than $^7$Li in SBBN model with the Planck baryon density.
The $^7$Be nuclei are then converted to $^7$Li nuclei via recombination with 
electron followed by the electron capture, i.e., $^7$Be + e$^-\rightarrow ^7$Li +$ \nu_e$.  
Therefore, the Li problem is alleviated if some exotic processes could destroy $^7$Be.  
One of solutions to the Li problem is an injection of nonthermal photon with energy of 
$\sim 2$ MeV which can destroy $^7$Be but not Deuterium (D) as calculated 
in Ref. \cite{Kusakabe:2013sna}.  If a long-lived exotic particle radiatively decays after BBN, nonthermal photons can disintegrate background thermal nuclei, and light 
element abundances change \cite{Lindley1979MNRAS.188P..15L}.  If the energy of the 
photon emitted at the decay is much larger than ${\cal O}(10)$ MeV, all of light nuclei are 
disintegrated by nonthermal photons~\cite{Lindley1979MNRAS.188P..15L,Ellis:1984er,
Dimopoulos:1987fz,1992NuPhB.373..399E,
Kawasaki:1994af,Kawasaki:1994sc,Jedamzik:1999di,Kawasaki:2000qr,Cyburt:2002uv,
Kawasaki:2004qu,Ellis:2005ii,Jedamzik:2006xz,Kusakabe:2006hc}.  In this case, therefore, 
the Li problem can not be solved (e.g., \cite{Ellis:2005ii,Kusakabe:2006hc}).  
Therefore, the energy of emitted photon for the $^7$Be destruction is limited 
to a narrow range \cite{Kusakabe:2013sna}.
A similar $^7$Be destruction would occur if a long-lived sterile neutrino 
decays into energetic electron and positron.  We then study cosmological effects of this decay channel in this paper.

From the theoretical point of view of the extended Minimal Standard Model
(MSM) of particle physics,
right-handed neutrinos introduced as sterile neutrinos provide an elegant
mechanism for
the generation of tiny active neutrino masses; so-called canonical seesaw
mechanism \cite{Minkowski:1977sc,Yanagida1979,Yanagida:1980xy,Gell-Mann1979}. 
If their masses are so heavy (more than $10^9~{\rm GeV}$), they also
explain the origin of the baryon asymmetry of the universe; so-called
leptogenesis scenario \cite{Fukugita:1986hr}.
Even if the sterile neutrinos have masses below electroweak (EW) scale,
however,
there exist other phenomenological effects without lacking the success
of the seesaw mechanism.
Several possibilities have been investigated regarding a detectability
of the sterile neutrinos, and
the case in which sterile neutrinos are lighter than light
mesons (e.g., pion or kaon) has been studied especially in detail \cite{Asaka:2011pb} (for
recent study, see \cite{Asaka:2012bb} and references therein).
The allowed sterile neutrino masses are smaller than the pion mass
$\sim$ 140 MeV, which are not consistent with recent results of
neutrino oscillation experiments,
unless one assumes that their lifetimes are longer than  $\sim 0.1$ s. 
In addition, another constraint has been derived from a study of BBN
by a comparison between theoretical and observational abundances of
$^4$He.
The upper limit on the lifetime of $\sim 0.1$ s was derived when a
relic abundance
of sterile neutrino is fixed as given in Ref.
\cite{Dolgov:2000pj,Dolgov:2000jw}.
However, this constraint depends on the relic abundance.
In this paper, we take into account the possibility that the abundance
is smaller
than the simple estimate \cite{Dolgov:2000pj,Dolgov:2000jw}.
In this case a longer lifetime of the sterile neutrino is allowed.

In this paper, we comprehensively investigate cosmological effects of a long-lived sterile neutrino with mass $\sim 10$ MeV.
In Sec. \ref{sec2}, we assume a decay of a sterile neutrino in the early universe, and describe our calculation method, and formulations of (1) the spectra of electrons and positrons generated at the decay, (2) those of primary photons induced by the energetic electrons and positrons, and (3) the nonthermal nucleosynthesis triggered by the energetic photons.  In Sec. \ref{sec3}, revised cross sections for photodisintegration of $^7$Be and $^7$Li are described.
In Sec. \ref{sec4}, effects of the decaying sterile neutrino on the cosmic thermal history, the effective neutrino number, and the baryon-to-photon ratios are formulated.
In Sec. \ref{sec5}, observational constraints on primordial light element abundances, the effective neutrino number, and the baryon-to-photon ratio adopted in this paper are summarized.
In Sec. \ref{sec6}, 
we show calculated energy spectra of electrons and positrons emitted at the decay, energy spectra of photons produced via the inverse Compton scattering of background photons by the electron and positron, and photon injection spectra resulting from electromagnetic cascade showers.  
Time evolutions of light element abundances, the baryon-to-photon ratio, and thermal and nonthermal neutrino energy densities are then consistently calculated with nonthermal photodisintegrations of nuclei taken into account.  An impact of revised cross sections of $^7$Be and $^7$Li photodisintegrations is also shown.
In Sec. \ref{sec7}, we discuss the relic abundance of the sterile neutrino before its decay.  We also comment on a possible dilution of the sterile neutrino in the early universe, effects of the sterile neutrino mixing to active neutrinos of different flavors, and experimental constraints from the pion decay and the supernova luminosity.  
In Sec. \ref{sec8}, this study is summarized.  In Appendix \ref{app1}, 
extensive formulae of the sterile neutrino decay are derived.  
We adopt natural units of $\hbar=c=k_{\rm B}=1$, where $\hbar$ is the reduced Planck constant, $c$ is the speed of light, and $k_{\rm B}$ is the Boltzmann constant.  
We also adopt notation of $A$($a$, $b$)$B$ for a reaction $A+a \rightarrow b + B$.

\section{Model}\label{sec2}

We simply include one right-handed (sterile) neutrino, $\nuh$, in the MSM 
and assume that it has a mass of ${\cal O}$(10) MeV and a rather long lifetime $\sim{\cal O}(10^5)$ s.
We consider effects of the decays of $\mathcal{O}(10) {\rm MeV}$ sterile neutrino on cosmological quantities, especially the $^7$Li number abundance relative to hydrogen, i.e., ${}^7 {\rm Li}$/H.

\subsection{Method}\label{sec2a}
We perform a BBN calculation.  Kawano's BBN code~\cite{Kawano1992,Smith:1992yy} is utilized
with default settings of the time steps in order to make the thermal nucleosynthesis calculation part as simple as possible.  To calculated results, we added Sarkar's correction for $^4$He abundances from explicit integration of weak rates, smaller time steps, Coulomb, radiative, and finite temperature corrections, and the correction for finite nucleon mass~\cite{Sarkar:1995dd}.  
Reaction rates for light nuclei ($A \le  10$) are updated with recommendations 
by JINA REACLIB Database V1.0 \cite{Cyburt2010}.  The neutron lifetime is set to be $880.0 \pm 0.9$~s from the weighted average value of the Particle data group~\cite{Beringer:1900zz}.  We note that after the improved measurements \cite{Serebrov:2004zf,Serebrov:2010sg,Mathews:2004kc}, a few earlier measurements have been reanalyzed, and updated lifetimes are significantly shorter than the previous ones~\cite{Beringer:1900zz}.

We adopt cosmological parameters reported from the analysis of the 
Planck~\cite{Ade:2013zuv}.  
Central values for the base $\Lambda$CDM model (Planck+WP+highL+BAO) determined from the Planck 2013 data are taken:  $H_0=67.3 \pm 1.2$ km s$^{-1}$ Mpc$^{-1}$, $\Omega_\Lambda=0.685 ^{+0.018}_{-0.016}$, $\Omega_{\rm m}=0.315 ^{+0.016}_{-0.018}$, and $\Omega_{\rm b} h^2=0.02205 \pm 0.00028$ with $h=H_0/$(100 km s$^{-1}$ Mpc$^{-1}$).  The baryon-to-photon ratio is calculated using most recent values of physical constants as follows.

\subsubsection{baryon-to-photon ratio}\label{sec2a1}
The present number density of CMB is given by
\begin{equation}
n_{\gamma 0} = \frac{2 \zeta(3)}{\pi^2} T_{\gamma 0}^3,
\label{n_g0}
\end{equation}
where $\zeta(3)=1.20205$ is the Riemann zeta function, and $T_{\gamma 0}$ is the present CMB temperature.  The present energy density of baryon is related to cosmological parameters as
\begin{equation}
\rho_{{\rm b} 0} = \Omega_{\rm b} \left( \frac{3 H_0^2}{8 \pi G_{\rm N}}\right),
\label{rho_b0}
\end{equation}
where $G_{\rm N}$ is the gravitational constant.  The present average mass per baryon is given \cite{Steigman:2006nf} by
\begin{eqnarray}
m_{{\rm b}0} &\equiv& \frac{\rho_{{\rm b}0}}{n_{{\rm b}0}} \nonumber \\
&=& \left\{ 1 -\left[ 1- \frac{1}{4} \left(\frac{m_{\rm He}}{m_{\rm H}}\right) Y_{\rm p}\right]\right\} m_{\rm H}\nonumber \\
&=& \left(1 -0.007119 Y_{\rm p} \right) m_{\rm H},
\label{m_b0}
\end{eqnarray}
where $n_{{\rm b}0}$ is the present number density of baryon, $m_{\rm H}=1.67353\times 10^{-21}$ and $m_{\rm He}=6.64648\times 10^{-21}$kg are the atomic masses of $^1$H and $^4$He, respectively \cite{Audi2003}, and $Y_{\rm p}$ is the cosmological mass fraction of $^4$He.  In the above equation, contributions from small abundances of D, $^3$He, $^7$Li, and heavier nuclides have been neglected.  

Using the above three equations, the relation between the baryon-to-photon ratio and the baryon density parameter is given \cite{Steigman:2006nf} by
\begin{eqnarray}
\frac{\eta}{\Omega_{\rm b} h^2} &=& \frac{n_{{\rm b}0}}{n_{\gamma0}} \left[\frac{3 \left( 100~{\rm km~s}^{-1}~{\rm Mpc}^{-1}\right)^2} {8 \pi G_{\rm N} m_{\rm H}} \right] \frac{m_{\rm H}}{\rho_{{\rm b}0}} \nonumber\\
&=& \frac{1}{n_{\gamma0}} \left[\frac{3 \left( 100~{\rm km~s}^{-1}~{\rm Mpc}^{-1}\right)^2} {8 \pi G_{\rm N} m_{\rm H}} \right] \frac{m_{\rm H}}{m_{{\rm b}0}} \nonumber\\
&=& 2.7378 \times 10^{-8} \left(\frac{G_{\rm N}}{6.6738\times 10^{-11}~{\rm m}^3~{\rm kg}^{-1}~{\rm s}^{-2}}\right)^{-1} \left( \frac{T_{\gamma 0}}{2.7255~{\rm K}}\right)^{-3} \left[ 1 +0.007131 \left( Y_{\rm p} -0.25\right) \right].
\label{eta_omegab_relation}
\end{eqnarray}

We adopt the CMB temperature \cite{Fixsen:2009ug}
\begin{eqnarray}
T_{\gamma 0} &=& 2.72548 \pm 0.00057~{\rm K} \nonumber\\
&=& 2.72548 (1 \pm 0.00021)~{\rm K}~~~~(1 \sigma).
\label{eq_Tcmb}
\end{eqnarray}
The latest gravitational constant \cite{Beringer:1900zz} is given by
\begin{eqnarray}
G_{\rm N} &=& \left( 6.67384 \pm 0.00080 \right)\times 10^{-11}~{\rm m}^3~{\rm kg}^{-1}~{\rm s}^{-2} \nonumber \\
&=& 6.67384 (1 \pm 0.00012) \times 10^{-11}~{\rm m}^3~{\rm kg}^{-1}~{\rm s}^{-2}~~~~(1 \sigma).
\label{eq_grav}
\end{eqnarray}
The square-root of sum of squares of uncertainties from $G_{\rm N}$ and $T_\gamma$ is taken, and the precise relation between $\eta$ and $\Omega_{\rm b}h^2$ is derived, for $Y_{\rm p} \sim 0.25$, as
\begin{equation}
\eta = (6.037 \pm 0.077) \times 10^{-10}~~~~(1 \sigma).
\end{equation}

\subsubsection{parameters}\label{sec2a2}
It is assumed that energetic electrons and positrons are generated at the decay of a 
long-lived massive particle (sterile neutrino) $\nuh$.  The sterile neutrino $\nuh$ has a mass $M_\nuh$ 
and a mean lifetime $\tau_\nuh$.   Through the inverse Compton scattering of cosmic background 
radiation (CBR), the energetic electrons and positrons produce energetic primary photons.  
The energies of the photons are related to the energy of $e^\pm$ and temperature of the universe.  

The present model has three parameters regarding effects of nonthermal nucleosynthesis: 
(1) $(n_\nuh^0/n_\gamma^0)$ is the number ratio of the decaying particle $\nuh$ and the background 
photon evaluated at a time between the cosmological electron positron pair annihilation and
the $\nuh$ decay, (2) the $\nuh$ mass $M_\nuh$, and (3) the lifetime $\tau_\nuh$.  The total energy of $e^\pm$ emitted at one $\nuh$ decay event $E_{\nuh\rightarrow e}$ is derived as a function of $M_\nuh$.  This quantity is equivalent to the energy injected in the form of electromagnetic cascade showers.  
We adopt the method of Ref.~\cite{Kusakabe:2006hc} to calculate the nonthermal nucleosynthesis, 
where thermonuclear reactions are simultaneously solved.  We utilized updated reaction rates of 
$^4$He photodisintegration (Eqs. (2) and (3) of Ref. \cite{Kusakabe:2008kf}) based on cross section 
data from measurements with laser-Compton photons~\cite{Shima:2005ix,Kii:2005rk}.  
In this work, we found errors in cross sections of reactions $^7$Be($\gamma$, $\alpha$)$^3$He 
and $^7$Li($\gamma$, $\alpha$)$^3$H \cite{Cyburt:2002uv} adopted in previous studies on the 
BBN model with the long-lived decaying particle.  
The errors are corrected in this calculation as explained in Sec. \ref{sec6}.

\subsection{injection spectrum of photon}\label{sec2b}
The injection spectrum of nonthermal photon is given by
\begin{equation}
p_\gamma(E_\gamma; T, M_\nuh) = \frac{1}{\Gamma} \int_{m_e}^{M_\nuh/2} \frac{d \Gamma}{d E_e}
\left(M_\nuh \right)~d E_e \int_0^{E_{\gamma0,{\rm max}}} P_{\rm iC} \left(E_e, E_{\gamma0}; T \right)~
p_{\gamma,{\rm EC}} \left(E_\gamma, E_{\gamma0}; T \right)~dE_{\gamma0},
\label{eq_inj}
\end{equation}
where
$T$ is the photon temperature of the universe, 
$m_e$ is the electron mass,
$(1/\Gamma)d \Gamma/dE_e(M_\nuh)$ is the differential decay rate as a function of energy 
of $e^\pm$ and $M_\nuh$ [cf. Eq. (\ref{eq_a15})],
$P_{\rm iC} \left(E_e, E_{\gamma0}; T \right)$ is the energy spectrum of primary photon ($E_{\gamma 0}$) 
produced at inverse Compton scatterings of $e^\pm$ with energy $E_e$ at temperature $T$, 
$p_{\gamma,{\rm EC}}(E_\gamma, E_{\gamma0}; T)$ is the energy spectrum of nonthermal photon with energy $E_\gamma$
which is produced in the electromagnetic cascade showers triggered by primary photon with energy $E_{\gamma0}$ at $T$, and
$E_{\gamma0,{\rm max}}(E_e, T)$ is the maximum value of $E_{\gamma0}$ [Eq. (\ref{e_g0_max})].

\subsection{sterile neutrino decay}\label{sec2c}
Throughout this paper, 
we concentrate on the Dirac neutrino case.
We denote the mass of the heaviest state of neutrinos as $M_\nuh$ and 
the active-sterile mixing angle as $\Theta$.

The energy spectrum of $e^\pm$, i.e., $d\Gamma/d E_e$ is calculated (Appendix \ref{app1}), and input in Eq. (\ref{eq_inj}).  In this paper, we fix parameters of the sterile neutrino as one possible simple example case as follows.  
The sterile neutrino couples to charged and neutral currents of the electron flavor only.  
The strengths of the $\nuh$ coupling to the currents are given by $\Theta$ (for electron) and $0$ (for muon and tauon).

 We note that the investigation in this paper can be easily extended to the Majorana case.
The only difference between the two cases is that decay rates of a Majorana sterile neutrino are twice as large as those of a Dirac neutrino when those neutrinos have the same parameter values of $M_\nuh$ and $\Theta$.
The relevant Lagrangian of neutrino sector including one Majorana sterile neutrino is given by
\begin{eqnarray}
\mathcal{L} &=& 
\bar{\nuh} \gamma_\mu {\partial^\mu} \nuh - F_{\alpha} \bar{L}_\alpha H \nuh \
- \frac{M_\nuh}{2} \bar{\nuh}^C \nuh + {\rm h.c.},
\end{eqnarray}
where $L_\alpha$ and $H$ are lepton and Higgs doublets, respectively, 
$F_\alpha$ (for $\alpha = e\,,\mu\,,\tau$) is the Yukawa coupling constant, 
and $M_\nuh$ is the Majorana mass.
After the breaking of EW symmetry, the sterile neutrino is mixed with active neutrinos. 
The degree of mixing is characterized by the active-sterile mixing 
denoted as $\Theta \equiv F_\alpha \langle H \rangle/M_\nuh$, where $\langle H \rangle$ denotes the vacuum expectation value of the Higgs field.

The sterile neutrino can decay through this mixing into $3\nu$ or $\nu \ell^+ \ell^-$, where $\ell$ is a charged lepton species.
Inversely, if sterile neutrino is relatively light, it can be produced by the decay of light mesons.  For instance when $M_\nuh < m_\pi$ (pion mass) is satisfied, the pion decay can produce the sterile neutrino.
In this case, we can find a signal of sterile neutrino as a new peak in the energy spectrum of active neutrinos.

\subsection{primary photon spectrum}\label{sec2d}
We assume that energetic electrons and positrons, $e^\pm$'s, are injected at the $\nuh$ decay.  Energetic photons, $\gamma$'s, are then 
produced via the inverse-Compton scattering between the $e^\pm$'s and 
CBR ($e^{\pm}+\gamma_{\rm bg} \rightarrow e^{\pm}+\gamma$), where the subscript ``bg'' means background. The inverse-Compton process plays 
an important role in developments of electromagnetic cascade shower.  It distributes the energy of $e^\pm$ 
generated at the decay to multiple particles, i.e., $\gamma$ and $e^\pm$ 
in the thermal bath.  We assume that the $e^\pm$ in the initial state with energy $E_e$ reacts with CBR 
with energy $E_{\gamma{\rm b}}$, and a scattered photon in the final state has energy $E_{\gamma0}$.  
Number of collisions for an $e^\pm$ particle per unit time $dt$ and unit energy interval of photon in the 
final state $dE_{\gamma0}$ is then approximately given~\cite{jon68,Kawasaki:1994sc} 
\footnote{As for Eq. (51) in Ref. \cite{Kawasaki:1994sc}, the minus sign in the denominator of the third term on 
the right hand side is different from that of Eq. (9) in Ref. \cite{jon68}.  We adopt the latter plus sign.} by
\begin{equation}
\frac{d^2 N}{dt d E_{\gamma0}}
= \frac{2\pi r_e^2 m_e^2}{E_{\gamma{\rm b}} E_{e}^2} 
F(E_{\gamma0}, E_e; E_{\gamma{\rm b}}),
\label{eqa14}
\end{equation}
where
$r_e=\alpha/m_e$ is the classical radius of electron with the fine structure constant $\alpha$, 
and the function $F(E_{\gamma0}, E_e; E_{\gamma{\rm b}})$ is defined by
\begin{equation}
F(E_{\gamma0}, E_e; E_{\gamma{\rm b}})=\left\{
\begin{array}{ll}
2q \ln q + (1+2q)(1-q)+\frac{(\gamma_E q)^2}{2(1+\gamma_E q)}(1-q) & {\rm for}~0 < q \leq 1 \\
0 & {\rm otherwise}, \label{eqa16}\\
\end{array}
\right.
\end{equation}
where parameters are introduced as
\begin{equation}
\gamma_E=\frac{4E_{\gamma{\rm b}} E_e}{m_e^2},
\hspace{3.5em}
q=\frac{E_{\gamma0}}{\gamma_E(E_e-E_{\gamma0})}.
\label{eqa17}
\end{equation}
The energy spectrum of CBR in the early universe is almost completely described by a Planck distribution,
\begin{equation}
f_{\gamma{\rm b}}(E_{\gamma{\rm b}})=\frac{E_{\gamma{\rm b}}^2}{\pi^2}\frac{1}{\exp(E_{\gamma{\rm b}}/T)-1}.
\label{eqa3}
\end{equation}

The energy of photon in the final state has an upper limit \cite{jon68},
\begin{equation}
E_{\gamma0}\leq E_{\gamma0,{\rm max}}=\frac{4 E_{\gamma{\rm b}} E_e^2}{m_e^2 
\left(1+4 E_{\gamma{\rm b}} E_e/m_e^2 \right)}.
\label{e_g0_max}
\end{equation}
When the maximum photon energy is between the threshold energies for the photodisintegration of $^7$Be (1.59 MeV) and that of D (2.22 MeV), an effective destruction of $^7$Be is possible without destructions of other light nuclides \cite{Kusakabe:2013sna}.  The maximum energy should, therefore, be in this energy range, i.e., $E_{\gamma0,{\rm max}}\sim 2$ MeV.  Here we approximately take the average energy of CBR, $\bar{E}_{\gamma{\rm b}}=2.701T$, as the CBR energy.  The $e^\pm$ energy required for the generation of nonthermal 
photon with $E_{\gamma0} \sim 2$ MeV is then estimated to be
\begin{eqnarray}
E_e &^>_\sim & \frac{1}{2}\left(E_{\gamma0,{\rm max}}+m_e \sqrt[]{\mathstrut 
\frac{E_{\gamma0,{\rm max}}}{E_{\gamma{\rm b}}}}\right) \nonumber\\
&\sim& 7.5
~{\rm MeV}~~~({\rm for}~t=10^6~{\rm s}).
\end{eqnarray}

Generally, at inverse Compton scatterings of low energy CBRs by energetic $e^\pm$ in the early universe, 
small fractions of energies of $e^\pm$ are transferred to those of CBRs at respective scatterings, i.e., 
$E_{\gamma0}~^\ll_\sim~E_e$.  The energy spectrum of the primary photon produced at the inverse Compton scattering is, 
therefore, approximately proportional to the differential scattering rate as a function of $E_{\gamma0}$.  
The spectrum of the primary photon is then given by
\begin{equation}
P_{\rm iC} \left(E_e, E_{\gamma0}; T \right)= \frac{F\left(E_{\gamma0}, 
E_e; E_{\gamma{\rm b}}\right)}{\int F\left(E_{\gamma0}, E_e; E_{\gamma{\rm b}}\right)~dE_{\gamma0}}
=\frac{F\left(E_{\gamma0}, E_e; E_{\gamma{\rm b}}\right)}
{\frac{E_e}{\gamma_E} \left[\frac{8+9\gamma_E +\gamma_E^2}{2 \gamma_E} \ln\left( 1+ \gamma_E \right) 
-\frac{16+18 \gamma_E +\gamma_E^2}{4 \left(1+ \gamma_E \right)} +2 {\rm Li}_2\left( -\gamma_E \right)\right]},
\label{eq_p_iC}
\end{equation}
where
Li$_2(-x)$ is the dilogarithm.  The dilogarithm is calculated in our code using the published Algorithm 490 \cite{Ginsberg1975}.

\subsection{electromagnetic cascade}\label{sec2e}
The energetic primary photons interact with the CBR, and electromagnetic cascade showers composed of 
energetic photons and $e^\pm$ develop (e.g.,~\cite{Ellis:1984er,Kawasaki:1994sc}).  
One energetic photon can produce multiple particles of lower energies by continuous reactions of pair 
production at a collision with CBR ($\gamma+\gamma _{\rm bg}\rightarrow e^+ + e^-$) 
and the inverse Compton scattering of $e^\pm$ at a collision with CBR ($e^\pm + \gamma _{\rm bg}\rightarrow e^\pm + \gamma$).  
The nonthermal photons then obtain a quasi-static equilibrium spectrum~\cite{Kawasaki:1994sc,Protheroe:1994dt}.

When the energy of the primary photon $E_{\gamma0}$ is much larger than the threshold energy for photodisintegration 
of light nuclides, i.e., $E_{\gamma0} \gg 1$ MeV, the steady state energy spectrum of the nonthermal photons is approximately 
given (e.g., \cite{berezinskii90,Cyburt:2002uv,Kusakabe:2006hc}) by
\begin{equation}
p_{\gamma,{\rm EC}}(E_\gamma,E_{\gamma0}; T)=\left\{
\begin{array}{ll}
K(E_X/E_\gamma)^{3/2} & {\rm for}~E_\gamma \leq E_X,\\
K(E_X/E_\gamma)^2 & {\rm for}~E_X < E_\gamma \leq E_C,\\
0 & {\rm for}~E_C < E_\gamma,\\
\end{array}
\right.
\label{eq_spe}
\end{equation}
where 
$E_X\sim m_e^2/(80T)$ and $E_C\sim m_e^2/(22T)$~\cite{Kawasaki:1994sc} are the energy corresponding to a 
break in the power law, and a cutoff energy, respectively, $K=E_{\gamma0}/\{E_X^2[2+\ln(E_C/E_X)]\}$ is the normalization 
constant which conserves the energy of the primary photon.  If nonthermal photons have energies larger than $E_C$, 
they are quickly destroyed via the electron-positron pair production. 

The maximum energy of the nonthermal photon $E_{\gamma0}$, however, can be of the order of $\mathcal{O}$(1 MeV) depending on the mass $M_\nuh$ and the temperature $T$ [Eq. (\ref{e_g0_max})].  
In this case, the generalized photon spectrum is given \cite{Kusakabe:2013sna} as follows:
(1)  For $E_{\gamma0} \leq E_X$, the spectrum is given by 
\begin{equation}
p_{\gamma,{\rm EC}1}(E_\gamma,E_{\gamma0}; T)=\left\{
\begin{array}{ll}
K_1(E_X/E_\gamma)^{3/2} & {\rm for}~E_\gamma \leq E_{\gamma0},\\
0 & {\rm for}~E_{\gamma0} < E_\gamma,\\
\end{array}
\right.
\end{equation}
where
$K_1=E_{\gamma0}^{1/2}/(2 E_X^{3/2})$.
\\
(2)  For $E_X< E_{\gamma0} \leq E_C$, the spectrum is given by
\begin{equation}
p_{\gamma,{\rm EC}2}(E_\gamma,E_{\gamma0}; T)=\left\{
\begin{array}{ll}
K_2(E_X/E_\gamma)^{3/2} & {\rm for}~E_\gamma \leq E_X,\\
K_2(E_X/E_\gamma)^2 & {\rm for}~E_X < E_\gamma \leq E_{\gamma0},\\
0 & {\rm for}~E_{\gamma0} <E_\gamma,\\
\end{array}
\right.
\end{equation}
where
$K_2=E_{\gamma0}/\{E_X^2[2+\ln(E_{\gamma 0}/E_X)]\}$.
\\
(3)  For $E_C< E_{\gamma0}$, the spectrum $p_{\gamma,{\rm EC}3}(E_\gamma,E_{\gamma0}, T)$ is given by Eq. (\ref{eq_spe}).

\subsection{steady state spectrum}\label{sec2f}

Rates of electromagnetic interactions are faster than the cosmic expansion rate.  The injection spectrum 
$p_\gamma(E_\gamma,T,M_\nuh)$ is then quickly modified to a new quasi-static equilibrium spectrum given by 
\begin{equation}
{\mathcal N}_\gamma^{\rm QSE}(E_\gamma; T, M_\nuh, \tau_\nuh, \zeta_{\nuh\rightarrow e}) =
\frac{n_\nuh(T; M_\nuh, \tau_\nuh, \zeta_{\nuh\rightarrow e}) p_\gamma (E_\gamma; T, M_\nuh)}{\Gamma_\gamma(E_\gamma; T, M_\nuh, \tau_\nuh, \zeta_{\nuh\rightarrow e})\tau_\nuh},
\label{ng}
\end{equation}
where 
\begin{equation}
n_\nuh(T; M_\nuh, \tau_\nuh, \zeta_{\nuh\rightarrow e})=n_\nuh^0(1+z)^3 \exp(-t(T; M_\nuh, \tau_\nuh, \zeta_{\nuh\rightarrow e})/\tau_\nuh)
\label{n_nu_h}
\end{equation}
is the number density of the decaying particles $\nuh$ at a redshift $z$, and $\zeta_{\nuh\rightarrow e}=(n_\nuh^0/n_\gamma^0)E_{\nuh\rightarrow e}$ is a parameter describing the amount of electromagnetic energy injection, with $n_\nuh^0$ and $n_\gamma^0$ the comoving number densities of $\nuh$ and CBR, respectively, estimated at a time between the cosmological $e^\pm$ annihilation and the $\nuh$ decay.  
Here, the cosmic time is described as $t(T; M_\nuh, \tau_\nuh, \zeta_{\nuh\rightarrow e})$ taking the inverse function of the temperature evolution $T(t; M_\nuh, \tau_\nuh, \zeta_{\nuh\rightarrow e})$.
The quantity $\Gamma_\gamma$ is the energy degradation rate of the nonthermal photons through 
three relatively slow processes; Compton scattering 
($\gamma + e^\pm_{\rm bg} \rightarrow\gamma + e^\pm$), Bethe-Heitler ordinary 
pair creation in nuclei ($\gamma + N_{\rm bg}\rightarrow e^+ + e^- + N$), 
and double photon scattering 
($\gamma + \gamma_{\rm bg} \rightarrow \gamma + \gamma$), where $N$ is a nucleon in nucleus.  Since the energy degradation rate depends not only on the photon temperature but also on the baryon and electron densities \cite{Kawasaki:1994sc}, it is a function of the baryon-to-photon ratio $\eta$.  We fix the $\eta$ value after the $\nuh$ decay to the observed value from the Planck CBR measurement.  The entropy production associated with the $\nuh$ decay changes the $\eta$ value as a function of time.  Therefore, the time evolution of the baryon-to-photon ratio depends on the parameters, i.e., $\eta=\eta(T; M_\nuh, \tau_\nuh, \zeta_{\nuh\rightarrow e})$.  We use this steady state approximation 
for the nonthermal photon spectrum. 

\subsection{nonthermal nucleosynthesis}\label{sec2g}
If the injection of nonthermal photons occurs at a cosmic time $t~^>_\sim~10^4$ s, the nonthermal photons 
can disintegrate background nuclei, and nuclear abundances can be changed~\cite{Lindley1979MNRAS.188P..15L,
Ellis:1984er,Dimopoulos:1987fz,1992NuPhB.373..399E,Kawasaki:1994af,Kawasaki:1994sc,Jedamzik:1999di,
Kawasaki:2000qr,Cyburt:2002uv,Kawasaki:2004qu,Ellis:2005ii,Jedamzik:2006xz,Kusakabe:2006hc,Kusakabe:2012ds}.  

\subsubsection{photodisintegration}\label{sec2g1}
The equation describing a time evolution of nuclear abundance by primary reactions, i.e., photodisintegration, is given by 
\begin{equation}
\frac{d Y_A}{d t} = \sum_P N_{AC}\left[P\gamma\right]_A Y_T - \sum_P \left[A\gamma\right]_P Y_A,
\label{dydt2}
\end{equation}
where 
$Y_i\equiv n_i/n_B$ is the mole fraction of a nuclear 
species $i$ with $n_i$ and $n_B$ number densities of nuclide $i$ and 
total baryon, respectively \footnote{The electromagnetic energy injection, as considered in this paper, increases the comoving number density of background photon as a function of time.  Therefore, the photon number density does not simply scale as $\propto (1+z)^3$.  Equations (\ref{dydt2})-(\ref{tt'}) are exact for a general case including the energy injection.  Equations for $dY/dz$ in Ref. \cite{Cyburt:2002uv} and $dY/dt$ in Ref. \cite{Kusakabe:2006hc} are, on the other hand, not exact when the effect of the electromagnetic energy injection by the exotic particle decay is significant.}, and we define the reaction rate 
\begin{equation}
\left[P\gamma\right]_A =
\frac{\eta_\nuh(T; M_\nuh, \tau_\nuh, \zeta_{\nuh\rightarrow e}) E_{\nuh \rightarrow e} }{\tau_\nuh} ~G_{[P\gamma]_A}^1(T; M_\nuh, \tau_\nuh, \zeta_{\nuh\rightarrow e}),
\label{nonthermal_rate1}
\end{equation}
where 
$\eta_\nuh=n_\nuh/n_\gamma$ is the $\nuh$-to-photon number ratio with $n_\gamma$ the number density of CBR, and the nuclear transfer function and the normalized spectrum of nonthermal photon are defined as
\begin{equation}
G_{[P\gamma]_A}^1(T; M_\nuh, \tau_\nuh, \zeta_{\nuh\rightarrow e}) =
\int_{0}^\infty dE_\gamma
T_\gamma^{\rm QSE}(E_\gamma; T, M_\nuh, \tau_\nuh, \zeta_{\nuh\rightarrow e})\,  \sigma_{\gamma +P \rightarrow A}
(E_\gamma),
\label{agamma}
\end{equation}
and
\begin{eqnarray}
T_\gamma^{\rm QSE}(E_\gamma; T, M_\nuh, \tau_\nuh, \zeta_{\nuh\rightarrow e})
&=&\frac{\tau_\nuh n_\gamma(T)}{E_ {\nuh \rightarrow e} n_\nuh(T; M_\nuh, \tau_\nuh, \zeta_{\nuh\rightarrow e})}{\mathcal N}_\gamma^{\rm QSE}(E_\gamma; T, M_\nuh, \tau_\nuh, \zeta_{\nuh\rightarrow e}) \nonumber \\
 &= &\frac{p_\gamma (E_\gamma, T, M_\nuh)}{E_{\nuh \rightarrow e} \left[ \Gamma_\gamma/n_\gamma \right](E_\gamma; T, M_\nuh, \tau_\nuh, \zeta_{\nuh\rightarrow e}) }.
 \label{sgamma}
\end{eqnarray}
The first and second term on the right hand side (RHS) in Eq. (\ref{dydt2}) correspond to the 
production ($\gamma+P\rightarrow A+C$) and destruction ($\gamma+A\rightarrow P+D$), respectively, for nuclide $A$.  
The cross section of the process $\gamma + P \rightarrow A+C$ is denoted by 
$\sigma_{\gamma + P \rightarrow A}(E_\gamma)$.  In addition, we use $N_{AC}$ as the number of identical nuclear species in a production process; $N_{AC}=2$ when particles $A$ and $C$ are identical 
and $N_{AC}=1$ when they are not.  For example, in the process 
$^4$He($\gamma$, $d$)D, $N_{\rm DD}=2$ since two deuterons are produced at one reaction.  

\subsubsection{secondary process}\label{sec2g2}
Nonthermal nuclei produced in the primary reaction can in general experience secondary nonthermal nuclear reactions.  
We, however, focus on an injection of photons with relatively small energies generated by a light sterile neutrino of $M_\nuh \leq 20$ MeV.  In this case, $^4$He photodisintegration is impossible, and most of important secondary reactions do not occur.  In this calculation, only the secondary reactions $^7$Li($\gamma$, $n$)$^6$Li($p$, $\alpha$)$^3$He and $^7$Be($\gamma$, $p$)$^6$Li($p$, $\alpha$)$^3$He are operative.
The equation describing the secondary production and destruction is given by 
\begin{equation}
\frac{d Y_S}{d t} = \sum_{P,A,P',X_1,X_2}Y_P Y_{P'} \frac{N_{AX_1} N_{SX_2}}{N_{AP'}}
 [P(A)P']_S -({\rm destruction~term}),
\label{dysdt}
\end{equation}
where the reaction rate for a secondary reaction $P(\gamma,X_1)A(P',X_2)S$ with any combination of particles 
$X_1$, $A$, and $X_2$ is given by
\begin{equation}
\left[P(A)P'\right]_S = 
\frac{\eta_\nuh(T; M_\nuh, \tau_\nuh, \zeta_{\nuh\rightarrow e}) E_{\nuh \rightarrow e} }{\tau_\nuh} 
~G_{[P(A)P']_S}^2(T; M_\nuh, \tau_\nuh, \zeta_{\nuh\rightarrow e}),
\label{nonthermal_rate2}
\end{equation}
\begin{eqnarray}
G_{[P(A)P']_S}^2(T; M_\nuh, \tau_\nuh, \zeta_{\nuh\rightarrow e}) &=&
\int_{0}^\infty dE_A \frac{\sigma_{A+P'\rightarrow S}(E_A)\beta_A(E_A)}
 {\left[ b_A/n_b \right](E_A; T,M_\nuh, \tau_\nuh, \zeta_{\nuh\rightarrow e})} \nonumber \\
& & \times \int_{{\mathcal E}_A^{-1}(E_A)}^\infty dE_\gamma
  T_\gamma^{\rm QSE} (E_\gamma; T, M_\nuh, \tau_\nuh, \zeta_{\nuh\rightarrow e}) \sigma_{\gamma + P\rightarrow A}
  (E_\gamma) \nonumber \\
 & & \times \exp{\left[ -\int_{E_A}^{{\mathcal E}_A(E_\gamma)} 
		   dE_A' \frac{\Gamma_A(E_A')}
		   {b_A(E_A'; T, M_\nuh, \tau_\nuh, \zeta_{\nuh\rightarrow e})}
		  \right] },
  \label{tt'}
\end{eqnarray}
where 
$\beta_A$ is the velocity of the nuclide produced in a primary reaction, i.e., primary nuclide, $A$, $b_A=-dE_A/dt$ is the energy loss rate of the primary nuclide mainly from Coulomb scattering of electrons, and
$\Gamma_A$ is the destruction rate of the primary nuclide.  Stable nuclides have $\Gamma_A=0$, while unstable nuclides have nonzero values given by the $\beta$-decay rates.  The quantity ${\mathcal E}_A(E_\gamma)$ is the
energy of the nuclide $A$ produced at the reaction $\gamma+P\rightarrow A$ of nonthermal photon with energy $E_\gamma$, and ${\mathcal E}_A^{-1}(E_A)$ is
the energy of the nonthermal photon which produces the primary nuclide $A$ with energy $E_A$.

The transfer functions $G_{[P\gamma]_A}^1(T; M_\nuh, \tau_\nuh, \zeta_{\nuh\rightarrow e})$ and $G_{[P(A)P']_S}^2(T; M_\nuh, \tau_\nuh, \zeta_{\nuh\rightarrow e})$ should be derived as a function of $T$ for a fixed parameter set of ($M_\nuh$, $\tau$, $\zeta_{\nuh\rightarrow e}$).  Time evolutions of the temperature $T(t)$, the baryon-to-photon ratio $\eta(t)$, and the $\nuh$-to-photon ratio $\eta_\nuh(t)=n_\nuh/n_\gamma$ are different for different parameter sets.  In Eq. (\ref{agamma}), the parameter dependence of the transfer function comes from the steady state nonthermal photon spectrum $p_\gamma(E_\gamma, T, M_\nuh)$ and the energy loss rate per background photon $[ \Gamma_\gamma /n_\gamma](E_\gamma; T, M_\nuh, \tau_\nuh, \zeta_{\nuh\rightarrow e})$.  In Eq. (\ref{tt'}), $b_A/n_b$ is independent of the baryon to photon ratio $\eta$ at the time since the energy loss rate of energetic nuclide, $b_A$, is proportional to the baryon density \cite{Reno:1987qw}.  However, it depends on the $\eta$ value at BBN through the $^4$He abundance (or the electron abundance).  Then, the quantity $b_A/n_b$, the photon spectrum $T_\gamma^{\rm QSE}$, and the energy loss rate $b_A$ in the exponential term depend on the parameter set ($M_\nuh, \tau_\nuh, \zeta_{\nuh\rightarrow e}$).

\subsubsection{approximation}\label{sec2g3}
In this study, we calculate the transfer functions neglecting the entropy production by the sterile neutrino decay.  Then, the $\eta$ value is constant, and set to be consistent with the central value determined by Planck.  Also the $\nuh$-to-photon ratio is exactly described as $\eta_\nuh=(n_\nuh^{\rm 0}/n_\gamma^{\rm 0})\exp(-t/\tau_\nuh)$ [cf. Eq. (\ref{n_nu_h})].  In this case, the transfer functions depend on the temperature $T$ and the mass $M_\nuh$ only.  The reaction rates [Eqs. (\ref{nonthermal_rate1}) and (\ref{nonthermal_rate2})] then reduce to
\begin{eqnarray}
\left[P\gamma\right]_A &=&
\frac{\zeta_{\nuh\rightarrow e}}{\tau_\nuh} \exp(-t/\tau_\nuh)~G_{[P\gamma]_A}^1(T; M_\nuh), \label{nonthermal_rate1b}
\\
\left[P(A)P'\right]_S &=& 
\frac{\zeta_{\nuh\rightarrow e}}{\tau_\nuh} \exp(-t/\tau_\nuh)~G_{[P(A)P']_S}^2(T; M_\nuh).
\label{nonthermal_rate2b}
\end{eqnarray}
We calculate nonthermal nucleosynthesis triggered by the nonthermal photons taking $\tau_\nuh$, $\zeta_{\nuh\rightarrow e}$, and $M_\nuh$ as parameters.  By using these simplified transfer functions, parameter search can be performed with same transfer functions for respective $M_\nuh$ values in realistic computation time.

We can safely use the simplified transfer functions without missing a parameter region in which the primordial $^7$Li abundance is reduced, for the following reason.  In this model, the entropy production by the $\nuh$ decay always reduces the baryon-to-photon ratio as a function of time (Sec. \ref{sec4}).  The ratio in the BBN epoch is, therefore, larger than that in the cosmological recombination epoch measured by Planck.  Then, one can place a lower limit on the BBN $\eta$ value taking the Planck value.   On the other hand, deuterium is only destroyed by photodisintegration, and is never produced in the nonthermal nucleosynthesis by the $\nuh$ decay.  The abundance of deuterium produced at the BBN, therefore, cannot be significantly smaller than the observational constraint on the primordial abundance.  This requirement gives an upper limit on the BBN $\eta$ value since primordial D abundance decreases as a function of $\eta$.  These limits are satisfied in a very narrow region of the BBN $\eta$ value, only ${\cal O}$(1) \% wide, which is around the Planck value (see fig. 1 of Ref. \cite{Kawasaki:2012va} or fig. 1 of Ref, \cite{Coc:2013eea}).  Then, maximum allowed changes of the $\eta$ value between the BBN and the recombination epochs is ${\cal O}$(1) \%.  Also the $\eta_\nuh$ value in the case with an entropy production can change from that in the case without it by only ${\cal O}$(1) \%.  The neglect of the entropy production effect in Eqs. (\ref{nonthermal_rate1b}) and (\ref{nonthermal_rate2b}), therefore, does not introduce a large error in final nuclear abundances.  In the following calculation, the final $\eta$ value is fixed to the Planck value.  When the entropy production changes the $\eta$ and $\eta_\nuh$ values by more than $\sim 10$ \%, present results of the nonthermal nucleosynthesis calculation are not precise.  However, such a large entropy production is accompanied by a large BBN $\eta$ value, and is therefore excluded by underproduction of deuterium definitely.

\section{photodisintegration cross sections of $^7$B\lowercase{e} and $^7$L\lowercase{i}}\label{sec3}
We correct significant errors in cross sections of reactions $^7$Be($\gamma$, $\alpha$)$^3$He and 
$^7$Li($\gamma$, $\alpha$)$^3$H \cite{Cyburt:2002uv} adopted in previous studies 
on the BBN model including a decaying particle (e.g., \cite{Kusakabe:2006hc,Kusakabe:2008kf,Kusakabe:2012ds,Kusakabe:2013sna}).  

The detailed balance relation between cross sections of a forward reaction $A$($B$, $\gamma$)$C$ and its inverse reaction $C$($\gamma$, $B$)$A$ is described \cite{Blatt} as 
\begin{equation}
\sigma_{C+\gamma}=\frac{g_A g_B}{(1+\delta_{AB}) g_C} \left(\frac{\mu E}{E_\gamma^2} \right) \sigma_{A+B},
\label{eq_b1}
\end{equation}
where
$\sigma_{A+B}$ and $\sigma_{C+\gamma}$ are the forward and inverse reaction cross sections, respectively,
$g_i=2I_i+1$ is the statistical degrees of freedom (DOF) with spin $I_i$ of species $i$,
$\delta_{AB}$ is the Kronecker delta for avoiding the double counting of identical particles,
$\mu$ and $E$ are the reduced mass and the center of mass (CM) energy, respectively, of the $A+B$ system, and
$E_\gamma=E+Q$ is the radiation energy with $Q$ the reaction $Q$-value: $Q=1.586627$ MeV 
[for $^3$He($\alpha$, $\gamma$)$^7$Be] and $Q=2.467032$ MeV [for $^3$H($\alpha$, $\gamma$)$^7$Li], respectively.

The forward reaction rate is described using the Astrophysical $S$-factor as
\begin{equation}
\sigma_{A+B}=\frac{S}{E}\exp \left( - \sqrt[]{\mathstrut \frac{E_{\rm G}}{E}} \right),
\label{eq_b2}
\end{equation}
where
$E_{\rm G}=(2\mu) (\pi Z_A Z_B \alpha)^2$ is the Gamow energy with
$Z_i$ the proton number of species $i$ and
$\alpha=1/137.04$ the fine structure constant.
Inserting this equation in Eq. (\ref{eq_b1}), we obtain a relation between the cross section of the inverse 
(photodisintegration) reaction and the $S$-factor of the forward (radiative capture) reaction:
\begin{equation}
\sigma_{C+\gamma}=\frac{g_A g_B}{(1+\delta_{AB}) g_C} \left(\frac{\mu}{E_\gamma^2} \right) S\exp \left( - \sqrt[]{\mathstrut \frac{E_{\rm G}}{E}} \right).
\label{eq_b3}
\end{equation}

Firstly, $S$-factors of the two reactions are taken from Ref. \cite{Kajino1987}.  When we take fitted functions 
[their Eqs. (6) and (7)] with theoretical values of $S(0)=0.511$ keV b for $^7$Be($\gamma$, $\alpha$)$^3$He 
and $S(0)=0.1003$ keV b for $^7$Li($\gamma$, $\alpha$)$^3$H, the photodisintegration cross sections are given by
\begin{eqnarray}
\sigma_{^7{\rm Be}+\gamma}&=& \frac{409~{\rm mb}}{E_{\gamma,{\rm MeV}}^2} 
\exp\left( -\frac{5.19}{E_{\rm MeV}^{1/2}} \right) \exp(-0.548 E_{\rm MeV}) \nonumber\\
&& \times \left(1-0.4285 E_{\rm MeV}^2 +0.5340 E_{\rm MeV}^3 -0.1150 E_{\rm MeV}^4 \right)~~~~~{\rm for}~Q 
\leq E_\gamma \leq Q+2.1~{\rm MeV},\nonumber \\
\label{eq_b4}
\end{eqnarray}
where
$E_{\gamma,{\rm MeV}}=E_\gamma/{\rm MeV}$ and $E_{\rm MeV}=E/{\rm MeV}$ are defined, and
\begin{eqnarray}
\sigma_{^7{\rm Li}+\gamma}&=& \frac{80.3~{\rm mb}}{E_{\gamma,{\rm MeV}}^2} 
\exp\left( -\frac{2.60}{E_{\rm MeV}^{1/2}} \right) \exp(-2.056 E_{\rm MeV}) \nonumber\\
&&\times \left(1 +2.2875 E_{\rm MeV}^2 -1.1798 E_{\rm MeV}^3 +2.5279 E_{\rm MeV}^4 \right)~~~~~{\rm for}~Q 
\leq E_\gamma \leq Q+1~{\rm MeV}.\nonumber \\
\label{eq_b5}
\end{eqnarray}
It is found that both cross sections are smaller than the published values \cite{Cyburt:2002uv} 
by a factor of about three although the values were based on the same reference \cite{Kajino1987}.  It is expected that this error originates from wrong treatment of 
statistical DOF.

Another error stems from the use of the fitted functions \cite{Kajino1987} which can be reasonably 
applied only to the low energy region of $E_\gamma-Q \lesssim {\cal O}(1)$ MeV.  
The functions have been derived from fitting to measured data in a relatively low energy region.  
These functions then provide erroneous values in a large $E_\gamma$ region.  
Especially, the function for $^7$Be($\gamma$, $\alpha$)$^3$He outputs large 
negative value at $E_\gamma \gtrsim 5.4$ MeV.  Since this error affects the final nuclear abundances 
calculated in the model with a decaying particle, it must be fixed.  For example, we may assume constant $S$-factors:  
$S=0.31$ keV b at $E_\gamma > Q+2.1$ MeV for $^7$Be($\gamma$, $\alpha$)$^3$He and $S=0.06$ keV b 
at $E_\gamma > Q+1$ MeV for $^7$Li($\gamma$, $\alpha$)$^3$H (cf. Ref. \cite{Angulo1999}).  
The cross sections are then given by
\begin{equation}
\sigma_{^7{\rm Be}+\gamma}= \frac{248~{\rm mb}}{E_{\gamma,{\rm MeV}}^2} 
\exp\left( -\frac{5.19}{E_{\rm MeV}^{1/2}} \right)~~~~~{\rm for}~Q+2.1~{\rm MeV} \leq E_\gamma,
\label{eq_b6}
\end{equation}
\begin{equation}
\sigma_{^7{\rm Li}+\gamma}= \frac{48.1~{\rm mb}}{E_{\gamma,{\rm MeV}}^2} 
\exp\left( -\frac{2.60}{E_{\rm MeV}^{1/2}} \right)~~~~~{\rm for}~Q+1~{\rm MeV} \leq E_\gamma.
\label{eq_b7}
\end{equation}

Recently, three independent groups have measured the cross section of the reaction $^3$He($\alpha$, $\gamma$)$^7$Be at high energies ($E \gtrsim 1.5$ MeV) with relatively small errors \cite{diLeva2009,diLeva2009b,Carmona-Gallardo2012,Bordeanu:2013coa}.  There is only one earlier publication for the cross section measured in this energy region, and the measured data included large errors \cite{Parker1963}.  The new measurements indicated a possible increase of the cross section at $E \gtrsim 1.5$ MeV which was not seen in the earlier work.  Theoretically, this increase can be understood as a contribution of the electric dipole capture from the scattering $d$-wave which gradually becomes more important at higher energies \cite{Tombrello1963,Liu1981}.  When we consider these experimental results, it may be better to take into account the $d$-wave behavior in an estimation of the cross section at high energies rather than to assume a constant $S$-factor.  We note, however, that a precise estimation needs more measurements at high energy of $E > 3$ MeV.  The only experimental data in the high energy region are from measurements for the $^3$He incident energy $E_3=19-26$ MeV at the angle 90$^\circ$ \cite{Waltham1983}.  The measured data typically give only upper limits on the differential cross section, $d \sigma/d \Omega(90^\circ) \lesssim 1 \mu$b/sr, at the corresponding CM energy $E\sim 12-17$ MeV.

An analytical function has been fitted to cross sections measured between 2004 and 2007 in the energy range of $0.04 \leq E \leq 1.2$ MeV \cite{Cyburt:2008up}.  Although this fitting did not take into account experimental data at higher energies, dominant contributions of $s$- and $d$-waves are included in the analytical function.  We then adopt this cross section function with six parameters in this paper.  The photodisintegration cross section is given by
\begin{eqnarray}
\sigma_{^7{\rm Be}+\gamma}&=& \frac{801~{\rm mb}}{E_{\gamma,{\rm MeV}}^2} 
\exp\left( -\frac{5.19}{E_{\rm MeV}^{1/2}} \right) \sum_{i=0,1} \frac{Q_i}{E+Q_i} \left[ s_{0i} \left( 1 +a_i E_{\rm MeV} \right)^2 +s_{2i} \left( 1+ 4\pi^2 \frac{E}{E_{\rm G}} \right) \left( 1 +16\pi^2 \frac{E}{E_{\rm G}} \right)\right],
\label{eq_b8}
\end{eqnarray}
where
$i=0$ and 1 indicate capture cross sections to the ground and the first excited states of $^7$Be, respectively,
and $Q_0=Q$ and $Q_1=1.1570$ MeV are the $Q$-values for the two final states.
The six fitted parameters are $s_{00}=0.406$, $s_{20}=0.007$, $a_0=-0.207$, $s_{01}=0.163$, $s_{21}=0.004$, and $a_1=-0.134$.

\begin{figure}
\begin{center}
\includegraphics[width=8.0cm,clip]{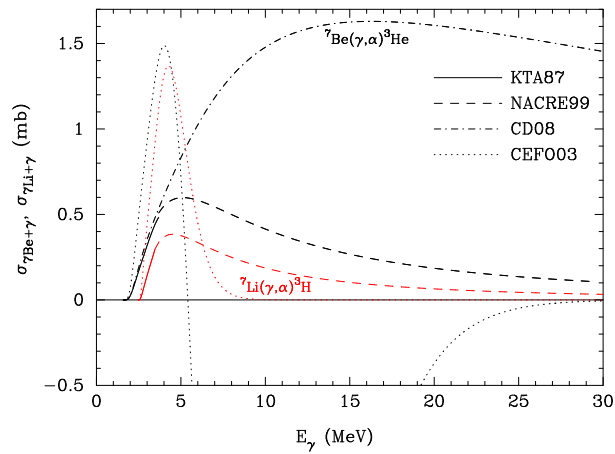}
\caption{Cross sections of reactions $^7$Be($\gamma$, $\alpha$)$^3$He and 
$^7$Li($\gamma$, $\alpha$)$^3$H as a function of the photon energy.  They are estimated from the detailed balance relation with the forward radiative capture cross sections.  Solid lines correspond to polynomial fits to theoretical calculations in the low energy regions \cite{Kajino1987}.  Dashed lines correspond to constant $S$-factors in the high energy regions \cite{Angulo1999}.  The dot-dashed line is from a fit to experimental data on $^3$He($\alpha$, $\gamma$)$^7$Be \cite{Cyburt:2008up}.  Dotted lines show fitted functions of Ref. \cite{Cyburt:2002uv}. \label{fig_a1}}
\end{center}
\end{figure}

Figure \ref{fig_a1} shows cross sections of reactions $^7$Be($\gamma$, $\alpha$)$^3$He and $^7$Li($\gamma$, $\alpha$)$^3$H as a function of the photon energy.  The cross sections are derived by applying the detailed balance relation to the forward radiative capture cross sections.  Solid and dashed lines correspond to polynomial fits to theoretical calculations in low energy regions \cite{Kajino1987} and constant $S$-factors in high energy regions \cite{Angulo1999}, respectively.  The dot-dashed line corresponds to fit to experimental data on $^3$He($\alpha$, $\gamma$)$^7$Be \cite{Cyburt:2008up}.  For comparison, dotted lines show fitted functions of Ref. \cite{Cyburt:2002uv}.   We adopt the dot-dashed line [Eq. (\ref{eq_b8})] for $^7$Be($\gamma$, $\alpha$)$^3$He, and the solid and the dashed lines [Eqs. (\ref{eq_b5}) and (\ref{eq_b7})] for $^7$Li($\gamma$, $\alpha$)$^3$H in the following calculations.


\section{effects on cosmic photon and neutrino background}\label{sec4}
The decay of sterile neutrino has two effects on resulting effective neutrino number in the universe, 
i.e., $N_{\rm eff}$.  Firstly, since the decay generates energetic neutrinos, the total energy density of neutrino 
increases with respect to the case of no neutrino injection \cite{Scherrer:1987rr,Fuller:2011qy}.  Secondly, since the decay also 
generates energetic electron and positron, the energy density of background photon increases \cite{Kolb:1981cx,Scherrer:1987rr,Fuller:2011qy}.  
The ratio between energy densities of neutrino and photon is reduced, and as a result,
effective neutrino number is reduced compared to the case of no photon injection.  The baryon-to-photon ratio is simultaneously reduced by this photon heating effect.  
In this study, we assume that the lifetime of the sterile neutrino is much larger than the time scale 
of the decoupling of the active neutrino in the early universe, i.e., $\tau_\nuh \gg 1$ s.  The energetic neutrino 
generated at the decay cannot interact effectively with background particles mainly constituted of weakly 
interacting neutrinos and weakly-noninteracting photon.  The energetic neutrino is then never thermalized, 
and propagates in the universe without collisions.  

\subsection{cosmic thermal history}\label{sec4a}
The total energy density in the universe is given \cite{Scherrer:1987rr} by
\begin{equation}
\rho =\rho_\gamma +\left( \rho_{e^-} + \rho_{e^+} \right) + \rho_{\nu,{\rm th}} +\rho_b +\rho_\nuh +\rho_{\nu,{\rm nt}},
\label{eq_rho_tot}
\end{equation}
where
$\rho_i$ is the energy density of particle species $i=\gamma$ (photon), $e^-$ (electron), $e^+$ (positron), $\nu,{\rm th}$ (thermal neutrino), $b$ (baryon), $\nuh$ (sterile neutrino), and $\nu,{\rm nt}$ (nonthermal neutrino produced at the $\nuh$ decay).  The last two terms are not present unless the decaying sterile neutrino exists.  The sterile neutrino energy density is given \cite{Scherrer:1987rr} by
\begin{eqnarray}
\rho_\nuh &=& M_\nuh n_\nuh \nonumber\\
          &=& M_\nuh n_\nuh^0 (1+z)^3 \exp \left(-t/\tau_\nuh \right) \nonumber \\
          &=& M_\nuh n_{\nuh,{\rm i}} \left( \frac{n_b}{n_{b,{\rm i}}} \right) \exp \left(-t/\tau_\nuh \right),
\end{eqnarray}
where
$n_i$ is the energy density of particle $i$, and
$n_{i,{\rm i}}$ is the energy density of $i$ at time $t_{\rm i}$ ($t_{\rm i} \gg 1$ MeV).  The initial number density of $\nuh$ is related as
\begin{equation}
n_{\nuh,{\rm i}} =\frac{11}{4} \frac{n_{\gamma,{\rm i}} \zeta_{\nuh \rightarrow e}}{E_{\nuh \rightarrow e}}.
\label{eq_n_nuh_i}
\end{equation}
The factor (11/4) originates from the entropy transfer from $e^\pm$ to photons at the cosmological $e^\pm$ annihilation.

When the lifetime of the sterile neutrino is equal to or longer than the BBN time scale, the cosmic expansion rate can be affected by the energy density of the sterile neutrino.  Since such a change in expansion rate changes resultant elemental abundances, it is constrained from observed light element abundances.  For example, the energy density of exotic relativistic species in the BBN epoch has been constrained \cite{Shvartsman1969,Steigman:1977kc}.  The sterile neutrino with a mass of ${\mathcal O}(10)$ MeV is nonrelativistic during BBN.  The energy density of a nonrelativistic particle redshift as $a^{-3}$, while that of a relativistic particle redshifts as $a^{-4}$.  Therefore, the former increases relative to the latter as the universe expands.  The effect of the nonrelativistic sterile neutrino is thus different from that of the exotic relativistic particle.  The change in the expansion rate during BBN is rigorously taken into account by using Eq. (\ref{eq_rho_tot}).

The nonthermal neutrino energy density is solved by time integration \cite{Scherrer:1987rr} of 
\begin{equation}
 \frac{d \rho_{\nu,{\rm nt}}}{dt} =-4H \rho_{\nu,{\rm nt}} + \frac{E_{\nuh\rightarrow \nu}}{M_\nuh} \frac{\rho_\nuh} {\tau_\nuh},
\label{drho_nudt}
\end{equation}
where 
$E_{\nuh\rightarrow \nu}$ is the average total energy of active neutrinos and antineutrinos emitted per one $\nuh$ decay event, and
the cosmic expansion rate is given by
\begin{equation}
H = \frac{\dot{a}}{a} = \left( \frac{8 \pi G_{\rm N}}{3} \rho \right)^{1/2},
\label{hubble}
\end{equation}
where $a$ is the scale factor of the universe.
The integral form of Eq. (\ref{drho_nudt}) is given by
\begin{eqnarray}
 \rho_{\nu,{\rm nt}}(t) &=& \frac{\rho_{\nuh,{\rm i}} a_{\rm i}^3}{a(t)^4} \frac{E_{\nuh\rightarrow \nu}}{M_\nuh} \frac{1}{\tau_\nuh} \int_{t_{\rm i}}^{t} a(t') {\rm e}^{-t'/\tau_\nuh} dt' \nonumber\\
&=& \left( \frac{11}{4} \right) n_{\gamma,{\rm i}} \frac{\zeta_{\nuh \rightarrow e}}{\tau_\nuh} \frac{a_{\rm i}^3}{a(t)^4} \int_{t_{\rm i}}^{t} a(t') {\rm e}^{-t'/\tau_\nuh} dt',
\label{eq_rho_nu_nt}
\end{eqnarray}
where $\rho_{\nuh,{\rm i}} \approx M_\nuh n_{\nuh,{\rm i}}$ is the initial energy density of $\nuh$, and
$a_{\rm i}$ is the scale factor at time $t_{\rm i}$.  We used Eq. (\ref{eq_n_nuh_i}) at the second equality.  

The ratio of the average total energy emitted in the form of electron and positron at the $\nuh$ decay to the mass $M_\nuh$ is given by
\begin{equation}
\frac{E_{\nuh\rightarrow e}}{M_\nuh} = \frac{1}{1+R(\nu, e)},
\end{equation}
where
$R(\nu, e)$ is the ratio of average total energies injected in the forms of $\nu$ (including all flavors and antineutrinos) and 
$e^\pm$ at the $\nuh$ decay.
In the present model, the sterile neutrino has only two decay modes, i.e., $\nuh\rightarrow \nu_e e^+ e^-$ and 
$\nuh\rightarrow \sum_{\beta =e, \mu, \tau} \nu_e \bar{\nu_\beta} \nu_\beta$.  The ratio of the decay rates for the two 
modes is defined with Eqs. (\ref{eq_a9}) and (\ref{eq_a21}) as
\begin{equation}
R_{\rm dec}=\frac{\Gamma (\nuh\rightarrow \nu_e e^+ e^-)}{\Gamma (\nuh\rightarrow \sum_\beta \nu_e \bar{\nu_\beta} \nu_\beta)}.
\label{eq_R_dec}
\end{equation}
Through the two decay modes, energetic neutrinos, electrons, and positrons are generated.  
The ratio of average total energies of $\nu$ and $e^\pm$ emitted through the two decay modes is given by
\begin{equation}
R(\nu, e)=\frac{1+R_{\rm dec} f_E(\nu_e)} {R_{\rm dec} \left[ f_E(e^-) +f_E(e^+) \right]},
\label{eq_ratio_nue}
\end{equation}
where
$f_E(i)=\bar{E_i}/M_\nuh$ is the ratio of the average energy of species $i$ to the sterile neutrino mass 
in the decay mode of $\nuh\rightarrow \nu_e e^+ e^-$, and the equation, $f_E(\nu_e)+f_E(e^-)+f_E(e^+)=1$ is satisfied.  
There is a trivial relation between parameters:
\begin{equation}
\frac{\zeta_{\nuh\rightarrow \nu}}{\zeta_{\nuh\rightarrow e}}=R(\nu, e),
\label{eq_zeta_ratio}
\end{equation}
where
$\zeta_{\nuh\rightarrow \nu}=(n_\nuh^0/n_\gamma^0)E_{\nuh\rightarrow \nu}$ is defined.

\begin{figure}
\begin{center}
\includegraphics[width=8.0cm,clip]{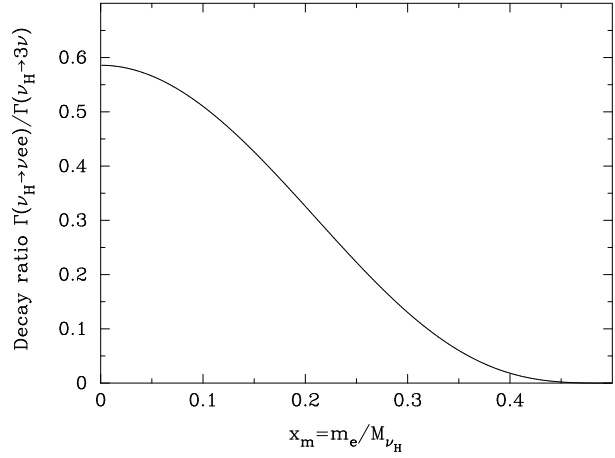}
\caption{The ratio of the decay rates $R_{\rm dec}=\Gamma (\nuh\rightarrow \nu_e e^+ e^-)/\Gamma (\nuh\rightarrow \sum_\beta \nu_e \bar{\nu_\beta} \nu_\beta)$ as a function of $x_{\rm m}=m_e/M_\nuh$. \label{fig8}}
\end{center}
\end{figure}

Figure \ref{fig8} shows the ratio of the decay rates $R_{\rm dec}$ as a function of $x_{\rm m}=m_e/M_\nuh$.  When the sterile neutrino mass is much larger than the electron mass, i.e., $x_{\rm m} \ll 0.5$, the rate for the decay into $\nu_e e^+e^-$ is comparable to that into three neutrinos.  The ratio monotonically decreases with increasing $x_{\rm m}$.  When the sterile neutrino mass is nearly one half of the electron mass, i.e., $x_{\rm m} \approx 0.5$, this ratio becomes very small and the decay into three neutrinos dominates.


\begin{figure}
\begin{center}
\includegraphics[width=8.0cm,clip]{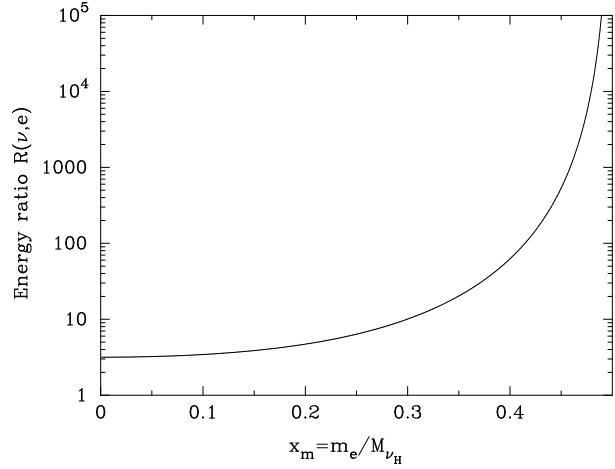}
\caption{The ratio of average total energies of neutrinos and $e^\pm$'s injected at the sterile neutrino decay $R(\nu, e)$ [Eq. (\ref{eq_ratio_nue})] as a function of $x_{\rm m}=m_e/M_\nuh$. \label{fig9}}
\end{center}
\end{figure}

Figure \ref{fig9} shows the ratio of average total energies of neutrinos and $e^\pm$'s injected at the sterile neutrino decay $R(\nu, e)$ as a function of $x_{\rm m}$.  In the large limit of the sterile neutrino mass, $x_{\rm m} \ll 1$, the ratio approaches to $\sim 10^{0.5}=3.16$.  The ratio monotonically increases with increasing $x_{\rm m}$.  The ratio diverges in the small mass limit, $x_{\rm m} \rightarrow \infty$.


Equation of the energy conservation is given (Eq. (D.26) of Ref. \cite{Kawano1992}) by
\begin{equation}
\frac{d} {dt} \left( \rho a^3 \right) + p \frac{d} {dt} \left( a^3 \right) + \left. a^3 \frac{d\rho}{dt}\right|_{T_9}^{\rm nuc} + \left. \frac{d(a^3 \rho_\nuh)}{dt}  \right|_{T_9}^{\rm dec} = 0,
\end{equation}
where $p$ is the total pressure of the universe, and 
the third and fourth terms in RHS correspond to the energy changes by nucleosynthesis and the electromagnetic energy injection at the sterile neutrino decay, respectively.  The term of the electromagnetic energy injection is given by
\begin{equation}
\left. \frac{d(a^3 \rho_\nuh)}{dt} \right|_{T_9}^{\rm dec} = - \frac{a^3 \rho_\nuh} {\tau_\nuh} \frac{E_{\nuh\rightarrow e}}{M_\nuh}.
\end{equation}
The equation of the energy conservation is then transformed to
\begin{equation}
\frac{dr}{d T_9} = -\frac{ \frac{d \rho_\gamma}{dT_9} + \frac{d \rho_e}{d T_9} + \frac{d \rho_b}{d T_9} } {\rho_\gamma + p_\gamma + \rho_e + p_e + p_b + \frac{1}{dr/dt} \left( \left. \frac{ d\rho_b}{dt} \right|_{T_9}^{\rm nuc} + \left. \frac{d \rho_e}{dt} \right|_{T_9}^{\rm nuc} -\frac{\rho_\nuh}{\tau_\nuh} \frac{E_{\nuh\rightarrow e}}{M_\nuh} \right)},
\label{drdt9}
\end{equation}
where
$r=\ln(a^3)$ is defined,
$dr/dt =3H$, and
$\rho_e =\rho_{e^-} + \rho_{e^+}$ and $p_e =p_{e^-} + p_{e^+}$ are the total energy and pressure, respectively, of electron and positron.  This is a varied form of the temperature evolution as a function of time, $dT_9/dt$ (cf. Eq. (22) of Ref. \cite{Scherrer:1987rr}), and is used in Kawano's BBN code.  The last term in the brackets in the denominator of RHS is absent unless the entropy production is induced by the $\nuh$ decay (cf. Eq. (D.28) of Ref. \cite{Kawano1992}).

\subsection{effective neutrino number}\label{sec4b}
The effective neutrino number is defined by
\begin{equation}
N_{\rm eff} =\frac{\rho_\nu}{\frac{7 \pi^2}{120} \left( \frac{4}{11} \right)^{4/3} T^4},
\label{eq_neff}
\end{equation}
where
$\rho_\nu= \rho_{\nu,{\rm th}} +\rho_{\nu,{\rm nt}}$ is the total neutrino energy density.  

The thermal neutrinos interact with thermal bath until the temperature decreases to $T \sim 1$ MeV.  Then, they decouple from background photons at $T\sim 1$ MeV, and have temperature $T_\nu$ which can be different from the photon temperature $T$.  The energy density of the thermal neutrinos is given by
\begin{equation}
\rho_{\nu,{\rm th}} = \frac{7 \pi^2}{120} N_{\rm eff,i} T_\nu^4,
\label{eq_rho_nu_th}
\end{equation}
where
$N_{\rm eff,i}=3$ is the effective neutrino number at the initial time $M_\nuh \gg t_{\rm i} > {\cal O}(1)$ MeV before the cosmological $e^\pm$ annihilation.  

The nonthermal neutrinos are assumed to originate only from the $\nuh$ decay.  Since we assume that the decay occurs much later than the BBN epoch, nonthermal neutrinos from the decay cannot be thermalized typically.  The nonthermal neutrinos, therefore, need to be treated separately from the thermal ones.  
The photon number density is given by
\begin{equation}
n_\gamma = \frac{2\zeta(3)}{\pi^2} T(t)^3.
\label{eq_rho_g}
\end{equation}
Using Eqs. (\ref{eq_rho_nu_nt}), (\ref{eq_neff}), (\ref{eq_rho_nu_th}), and (\ref{eq_rho_g}), the effective neutrino number is transformed to
\begin{equation}
N_{\rm eff} =\left( \frac{11}{4} \right)^{4/3} \left( \frac{T_\nu}{T} \right)^4 N_{\rm eff,i} +\frac{240 \zeta(3)}{7 \pi^4} \left( \frac{11}{4} \right)^{7/3} \frac{ \zeta_{\nuh \rightarrow \nu}}{\tau_\nuh}  \frac{T(t_{\rm i})^3}{T^4} \frac{a_i^3}{a(t)^4}\int_{t_{\rm i}}^t a(t') {\rm e}^{-t'/\tau_\nuh} dt'.
\end{equation}
We use $T(t_{\rm i})=T_\nu(t_{\rm i})$ and $a(t)T_\nu(t) =$constant.  The effective neutrino number can then be described as
\begin{equation}
N_{\rm eff} =\left( \frac{11}{4} \right)^{4/3} \left( \frac{T_\nu}{T} \right)^4 \left[ N_{\rm eff,i} +\frac{240 \zeta(3)}{7 \pi^4} \left( \frac{11}{4} \right) \int_{t_{\rm i}}^t \frac{\zeta_{\nuh\rightarrow \nu}}{T_\nu(t')} {\rm e}^{-t'/\tau_\nuh} \frac{dt'}{\tau_\nuh} \right].
\label{eq_neff_final}
\end{equation}
The factor $(11/4)^{4/3} (T_\nu/T)^4$ in RHS results from the entropy production by the injection of nonthermal electron and positron at the $\nuh$ decay.  The number and energy densities of thermal neutrinos for a fixed photon temperature $T$ are decreased 
compared to the case of no entropy production. 
The second term in the square brackets corresponds to an increase of the neutrino energy density contributed by the nonthermal active neutrino injection.  

The time derivative of the effective neutrino number can be derived from this equation as
\begin{equation}
\frac{d N_{\rm eff}}{dt} =4 N_{\rm eff} \left( \frac{d \ln T_\nu}{dt} -\frac{d \ln T}{dt} \right) +\left( \frac{11}{4} \right)^{7/3} \left( \frac{T_\nu}{T} \right)^4 \frac{240 \zeta(3)}{7 \pi^4} \frac{\zeta_{\nuh\rightarrow \nu}}{T_\nu} {\rm e}^{-t/\tau_\nuh} \frac{1}{\tau_\nuh}.
\end{equation}
Then, we solve the time evolution of the effective neutrino number simultaneously with those of $a(t)$ [Eq. (\ref{hubble})], $T(t)$ [Eq. (\ref{drdt9})], and $T_\nu(t) \propto a(t)^{-1}$.

\subsection{baryon-to-photon ratio}\label{sec4c}
An injection of energetic $e^\pm$ at the $\nuh$ decay produces nonthermal photons via electromagnetic cascade showers, and enhances the comoving photon entropy in the universe.  Since the baryon-to-photon ratio $\eta$ is inversely proportional to the comoving photon entropy, the ratio is reduced as a function of time during the nonthermal photon injection \cite{Kolb:1981cx,Feng:2003uy}.  In this case, the $\eta$ value in the BBN epoch is larger than that in the epoch of the CBR last scattering.  
We adopt the baryon-to-photon ratio inferred from Planck measurement of CBR as the value after the $\nuh$ decay.  The initial $\eta$ value is then determined as a function of $\tau_\nuh$ and $\zeta_{\nuh\rightarrow e}$ such that it results in the final $\eta$ value consistent with the Planck data.  Here we define the following variables:  The time $t_{\rm bef}$ is between the cosmological $e^\pm$ annihilation epoch and the $\nuh$ decay, while the time $t_{\rm aft}$ is after the decay.  The quantities $S_{\gamma,{\rm bef}}$ and $S_{\gamma,{\rm aft}}$ are the comoving photon entropies at times $t_{\rm bef}$ and $t_{\rm aft}$, respectively, while $\eta_{\rm bef}$ and $\eta_{\rm aft}$ are baryon-to-photon ratios at times $t_{\rm bef}$ and $t_{\rm aft}$, respectively.  

The comoving entropy density is given \cite{kolb1990} by
\begin{equation}
S=g_{\ast{\rm S}} a^3 T^3,
\end{equation}
where 
$g_{\ast{\rm S}}$ is the relativistic DOF in terms of entropy.  Since we consider the $\nuh$ decay after the neutrino decoupling at $T \sim 1$ MeV, the photon is the only relativistic component in the thermal bath.  Therefore, this DOF is the same as the statistical DOF of photon, i.e., $g_{\ast{\rm S}}=g_\gamma=2$.  The scale factor of the universe is inversely proportional to the neutrino temperature.  The ratio of the comoving photon entropies $S_{\gamma}/S_{\gamma,{\rm bef}}=\eta_{\rm bef}/\eta$ is, therefore, given by 
\begin{eqnarray}
\frac{S_{\gamma}}{S_{\gamma,{\rm bef}}} & =&\left(\frac{T_{\nu,{\rm bef}}}{T_{\rm bef}}\right)^3 \left(\frac{T}{T_{\nu}}\right)^3 \nonumber \\
&=& \left(\frac{4}{11}\right) \left(\frac{T}{T_\nu}\right)^3,
\end{eqnarray}
where
$T_{\nu,{\rm bef}}$ is the neutrino temperature before the $\nuh$ decay, and
we used the relation $T_{\nu,{\rm bef}}/T_{\rm bef}=(4/11)^{1/3}$.

The Planck measurement has obtained the $\Omega_b h^2$ value from combined data of Planck+WP+highL+BAO:  $\Omega_b h^2=0.02214\pm 0.00024$ (68\% C.L.) \cite{Ade:2013zuv}.  Therefore, the error in the baryon-to-photon ratio [Eq. (\ref{eta_omegab_relation})] or the comoving entropy at the CBR last scattering is lower than 2.2\% 
($2\sigma$).

\subsection{approximate formulae}\label{sec4d}
\subsubsection{baryon-to-photon ratio}\label{sec4d1}

When the comoving photon entropy changes by a small fraction, i.e., $\ll 100$ \% in the epoch of the $\nuh$ decay, the ratio $S_{\gamma,{\rm aft}}/S_{\gamma,{\rm bef}}=\eta_{\rm bef}/\eta_{\rm aft}$ is approximately given \cite{Feng:2003uy} by 
\begin{equation}
\frac{S_{\gamma,{\rm aft}}}{S_{\gamma,{\rm bef}}}=\exp\left[\frac{45^{3/4} 
\zeta(3)}{\pi^{11/4}}~\frac{\left(g_\ast^{\tau_\nuh}\right)^{1/4}}{g^{\rm bef}_{\gamma{\ast S}}} 
~\frac{E_{\nuh\rightarrow e}  n_\nuh^{\rm bef}}{n_\gamma^{\rm bef}} ~\sqrt{\frac{\tau_\nuh}{M_{\rm Pl}}}
\right],
\end{equation}
where 
$g_\ast^{\tau_\nuh}=3.36$ and $g_{\gamma{\ast {\rm S}}}^{\rm bef}=g_{\gamma}=2$ are
the relativistic DOFs in terms of energy and entropy, respectively, after the BBN epoch, 
$n_\nuh^{\rm bef}$ and $n_\gamma^{\rm bef}$ are number densities of the decaying particle and photon, respectively, 
evaluated at the same time $t_{\rm bef}$, and $M_{\rm Pl}=G_{\rm N}^{-1/2}=1.22\times 10^{19}$ GeV is the Planck mass.  The $g_{\gamma{\ast {\rm S}}}^{\rm bef}$ value is the same as statistical DOF of photon \footnote{We should take 
into account only the photon as a relativistic species in thermal and chemical contact with $e^\pm$ and 
photon itself during the cosmic epoch after BBN.  This is because light active neutrinos had been decoupled 
from thermal bath before the onset of BBN at $T \sim 1$ MeV.  We note that previous studies 
(e.g., \cite{Feng:2003uy,Kusakabe:2013sna}) are erroneous since the $g_{\ast S}$ factor included the contribution 
from neutrinos.  The right numerical value in the following Eq. (\ref{eq_entropy}) is then higher than the previous 
suggestion by a factor of 3.91/2=1.96.}.  For a small fractional change of entropy, 
this value is given \cite{Feng:2003uy} by
\begin{equation}
\frac{\Delta S_\gamma}{S_\gamma}\approx \ln \frac{S_{\gamma,{\rm aft}}}{S_{\gamma,{\rm bef}}}=2.14 \times 10^{-4}
\left(\frac{\zeta_{\nuh\rightarrow e}}{10^{-9}~{\rm GeV}}\right) \left(\frac{\tau_\nuh}{10^6~{\rm s}}\right)^{1/2}.
\label{eq_entropy}
\end{equation}
The ratio of the baryon-to-photon ratios $\eta_{\rm aft}/\eta_{\rm bef}=S_{\gamma,{\rm bef}}/S_{\gamma,{\rm aft}}$ can be estimated using this formula.

\subsubsection{neutrino number}\label{sec4d2}
The energy density of nonthermal active neutrino is generated by the $\nuh$ decay, and an approximate formula can be derived as in the case of a nonthermal photon injection in the late universe \cite{Hu:1993gc}.  It is assumed that the sterile neutrino does not contribute to the total energy density so much that the cosmic expansion rate is not affected much.  The standard radiation dominated universe then holds.  We define the second term in the square brackets of Eq. (\ref{eq_neff_final}) as $N_{\rm eff,nt,i}$.  This term then approximately becomes
\begin{eqnarray}
N_{\rm eff,nt,i} &=& \frac{240 \zeta(3)}{7 \pi^4} \left( \frac{11}{4} \right) \frac{\zeta_{\nuh\rightarrow \nu}}{T_\nu(t_{\rm eff})} \nonumber\\
&\approx& \frac{240 \zeta(3)}{7 \pi^4} \left( \frac{11}{4} \right)^{4/3} \frac{\zeta_{\nuh\rightarrow \nu}}{T(t_{\rm eff})},
\label{Neff_nt_i}
\end{eqnarray}
where
the effective time is defined as $t_{\rm eff}=[\Gamma(1+\beta)]^{1/\beta} \tau_\nuh$ for the universe with the 
time-temperature relation of $T\propto t^{-\beta}$ with $\Gamma(x)$ the gamma function of argument $x$ 
\footnote{We derived this effective time \cite{Kusakabe:2006hc} which is different from 
$t_{\rm eff}=[\Gamma(\beta)]^{1/\beta} \tau_\nuh$ in Ref. \cite{Hu:1993gc}.}.  
For the radiation dominated universe considered here, $\beta=1/2$ and $t_{\rm eff}=(\pi/4)\tau_\nuh$ are satisfied.

If the nonthermal neutrino injection is not accompanied by the nonthermal photon injection, the photon heating never occurs and the effective neutrino number after the decay is $N_{\rm eff}=N_{\rm eff,i} +\Delta N_{\rm eff,nt,i}$.  On the other hand, when nonthermal electrons and positrons are injected at the $\nuh$ decay, the entropy of the universe is increased.  
The number and energy densities of neutrino for a fixed photon temperature $T$ are then decreased 
compared to the case of no entropy production.  The effective number after the decay is given \cite{Fuller:2011qy} by
\begin{equation} 
N_{\rm eff} = \left( \frac{11}{4} \right)^{4/3} \left( \frac{T_\nu}{T}\right)^4 \left( N_{\rm eff,i} + N_{\rm eff,nt,i} \right)
= \left(\frac{S_{\gamma,{\rm aft}}}{S_{\gamma,{\rm bef}}}\right)^{-4/3} \left( N_{\rm eff,i} + N_{\rm eff,nt,i} \right).
\label{eq_dn}
\end{equation}
An approximate solution of $N_{\rm eff}$ is then derived with Eqs. (\ref{eq_entropy}), (\ref{Neff_nt_i}), and (\ref{eq_dn}).

\section{observational constraints}\label{sec5}
We constrain the model of the decaying sterile neutrino by comparing calculated results and observational constraints 
on elemental abundances, the effective neutrino number, and the CMB energy spectrum.  

\subsection{light element abundances}\label{sec5a}
In this model, $^7$Be is disintegrated by nonthermal photons originating from the $\nuh$ decay.  Since a primordial abundance of $^7$Li is mainly contributed from that of $^7$Be produced during 
the BBN epoch, the destruction of $^7$Be reduces the final $^7$Li abundance.  However, the energies of nonthermal 
photons should be small since energetic photons can disintegrate other nuclei and result in an inconsistency 
with observed abundances.  In the case of low energy photons, abundances of only D, $^7$Li, and $^7$Be can be significantly 
affected because of their small threshold energies for photodisintegration \cite{Kusakabe:2013sna}. If the photon energy is larger (4 MeV $<E_\gamma < 20$ MeV), the photodisintegration of $^6$Li, $^3$He and $^3$H, and the production of $^6$Li via $^7$Be($\gamma$, $p$)$^6$Li and $^7$Li($\gamma$, $n$)$^6$Li are also possible.  In addition, the time evolution of the baryon-to-photon ratio induced by the $\nuh$ decay changes light element abundances from the values in SBBN.  Therefore, we adopt the following constraints for respective light nuclides.  It is noted that only the constraints on D and $^7$Li abundances are important while other constraints are not in deriving constraints in the parameter plane of this model with the decaying sterile neutrino of $M_\nuh \sim 10-20$ MeV (Sec. \ref{sec6b}).

We use the primordial D abundance determined from observations of quasistellar object (QSO) absorption systems.  
As conservative constraints, the $2\sigma$ and $4\sigma$ ranges estimated with the mean value of ten Lyman-$\alpha$ absorption 
systems, log(D/H)$=-4.58\pm 0.02$ ($1\sigma$)~\cite{Pettini:2012ph}, are adopted.

$^3$He abundances are measured in Galactic \HII\ regions through the $^3$He$^+$ 8.665~GHz hyperfine transition line, $^3$He/H=$(1.9\pm 0.6)\times 10^{-5}$ ($1\sigma$)~\cite{Bania:2002yj}.  Since the uncertainty in the estimation of primordial $^3$He abundance is large, this constraint should be regarded only as a guide.  We take a $2\sigma$ upper limit from this observation.

Primordial $^4$He abundance has been derived by two different observations of metal-poor extragalactic \HII\ regions:  $Y_{\rm p}=0.2565\pm 0.0051$ ($1\sigma$)~\cite{Izotov:2010ca} and $Y_{\rm p}=0.2561\pm 0.0108$ ($1\sigma$)~\cite{Aver:2010wq}.  We take $2\sigma$ limits from the latter conservative result with a large error bar.

Primordial $^7$Li abundance is taken from a determination by spectroscopic observations of MPSs.  The observed abundances are about three times smaller than theoretical values in the SBBN model.  We adopt the 
observational limit, log($^7$Li/H)$=-12+(2.199\pm 0.086)$  ($1\sigma$) derived in a 3D nonlocal thermal equilibrium model~\cite{Sbordone:2010zi}.  We should consider the possibility that $^7$Li abundances in surfaces of MPSs are depleted by a factor of $\lesssim 2$ as suggested by calculations of a stellar model with a turbulent mixing \cite{Richard:2004pj,Korn:2007cx,Lind:2009ta}.  Below we find that the $^7$Li abundance in this model cannot be consistent with the observational $2\sigma$ limit.  When the $^7$Li depletion occurs in the stellar surfaces, however, some degree of $^7$Li reduction in this model can explain the observation.

$^6$Li abundances are measured in observations of MPSs.  A recent analysis, however, does not indicate any detection of this isotope \cite{Lind:2013iza}.  Since $^6$Li nuclei can be reasonably produced and destroyed after BBN, its primordial abundance is chosen conservatively.  We use the least stringent $2\sigma$ upper limit among those for all stars reported in Ref. \cite{Lind:2013iza}: $^6$Li/H=$(0.9\pm 4.3)\times 10^{-12}$ ($1\sigma$) for the G64-12 (nonlocal thermal equilibrium model with 5 free parameters).  

\subsection{effective neutrino number}\label{sec5b}
The effective neutrino number has been constrained from CMB observations.  We adopt the latest limit derived from the combined data of Planck+WP+highL+BAO for the one-parameter extension to the base $\Lambda$CDM model: $N_{\rm eff} =3.30^{+0.54}_{-0.51}$ (95\% C.L.) \cite{Ade:2013zuv}.  This one-parameter extension model is different from the present model of decaying sterile neutrino in terms of cosmic expansion rates and the time evolutions of $N_{\rm eff}$ values.  In the former model, relativistic particles are added to the standard $\Lambda$CDM model, while in the latter, the nonrelativistic $\nuh$ and its decay are assumed.  The above limit, therefore, cannot be applied to the present model as it is.  This limit, however, mainly comes from effects of the radiation energy density around the matter-radiation equal time.  Since the equal time is much later than the decay lifetime of the sterile neutrino $\tau_\nuh \lesssim 10^6$ s (see Sec. \ref{sec3}) considered in this paper, the above limit approximately gives a limit on the final $N_{\rm eff}$ value in this decaying $\nuh$ model.

\subsection{CMB energy spectrum}\label{sec5c}
An injection of nonthermal photons to the thermal bath triggers a deformation of the CBR spectrum from black-body \cite{Hu:1993gc}.  
Such a deformation is severely constrained by observations which indicate a nearly complete Planck spectrum 
\cite{Fixsen:1996nj,Seiffert2011}.  We adopt the most stringent and reliable limits on the CMB energy spectrum as follows.  
The limit on the chemical potential is taken from the analysis of the data from the Far-InfraRed Absolute Spectrophotometer 
on board the COsmic Background Explorer, $|\mu|<9\times 10^{-5}$~\cite{Fixsen:1996nj}.  The limit on the Compton $y$-parameter 
is taken from an updated constraint from the second generation of the Absolute Radiometer for Cosmology, Astrophysics, 
and Diffuse Emission (ARCADE) \footnote{http://arcade.gsfc.nasa.gov.} utilizing a better fitting procedure, 
$|y|<1\times 10^{-4}$~\cite{Seiffert2011}.  

\section{Results}\label{sec6}

\subsection{Nonthermal photon spectra}\label{sec6a}

\begin{figure}
\begin{center}
\includegraphics[width=8.0cm,clip]{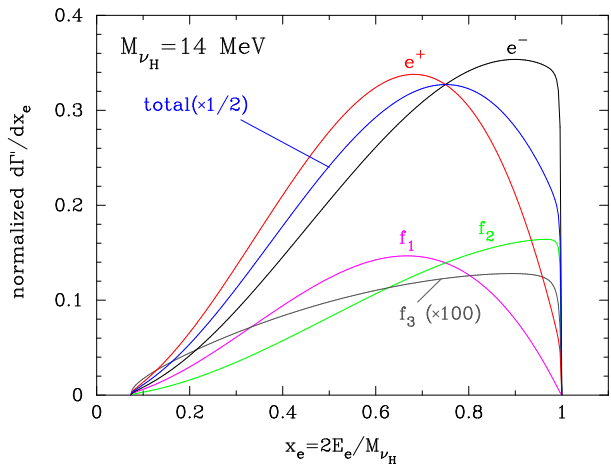}
\caption{ Normalized energy spectra of electron [Eq. (\ref{eq_a23})] and positron [Eq. (\ref{eq_a24})] generated at the $\nuh$ decay, 
and the total spectrum.  Functions $f_1(x_e)$, $f_2(x_e)$ and $f_3(x_e)$ [cf. Eqs. (\ref{eq_a25})-(\ref{eq_a27})] are also shown. 
The mass of the sterile neutrino is assumed to be $M_\nuh=14$ MeV. \label{fig4}}
\end{center}
\end{figure}

Figure \ref{fig4} shows normalized energy spectra of electron [Eq. (\ref{eq_a23})] and positron [Eq. (\ref{eq_a24})]
generated at the $\nuh$ decay, and the total spectrum (multiplied by 1/2) as a function of $x_e=2E_e/M_\nuh$ with $E_e$ the energies of electron and positron.  For all figures 
in this subsection, we assume that the mass of the sterile neutrino is $M_\nuh=14$ MeV since we find that this mass is included in the most important mass region for $^7$Be photodisintegration in the present model (see Sec. \ref{sec6b}).  
Functions $f_1(x_e)$, $f_2(x_e)$, and $f_3(x_e)$ [cf. Eqs. (\ref{eq_a25})-(\ref{eq_a27})] are also shown.  Because of these extended 
energy spectra of $e^\pm$, nonthermal photons produced via inverse Compton scattering of CBR by 
the generated $e^\pm$ also have extended energy spectra. 


\begin{figure}
\begin{center}
\includegraphics[width=8.0cm,clip]{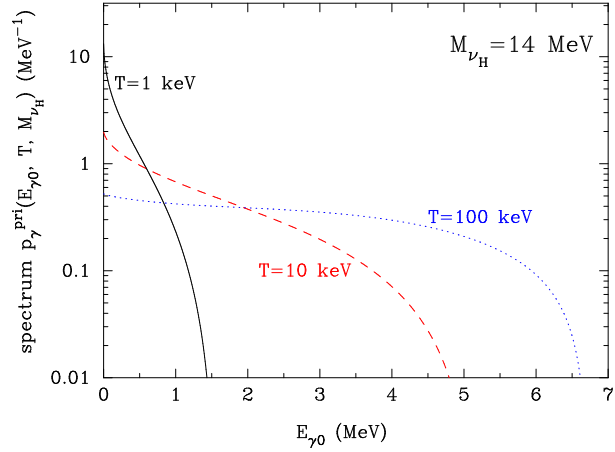}
\caption{Energy spectra of primary photon produced via the inverse Compton scattering of electron and positron 
which are generated at the decay for the cosmic photon temperature of $T=1$, 10 and 100 keV, respectively.  
The mass of the sterile neutrino is assumed to be $M_\nuh=14$ MeV. \label{fig5}}
\end{center}
\end{figure}

Figure \ref{fig5} shows energy spectra of primary photon produced via the inverse Compton scattering of electron 
and positron which are generated at the decay for the cosmic photon temperature of $T=1$, 10 and 100 keV, respectively.  
The primary photon spectra is given by
\begin{equation}
p_\gamma^{\rm pri}(E_{\gamma0}; T, M_\nuh) = \frac{1}{\Gamma} \int_{m_e}^{M_\nuh/2} \frac{d \Gamma}{d E_e}\left(M_\nuh \right) P_{\rm iC} \left(E_e, E_{\gamma0}; T \right)~d E_e.
\label{eq_pg_pri}
\end{equation}
When the energy injection with given spectra (Fig. \ref{fig4}) occurs at lower temperatures, the inverse Compton 
scattering produces softer spectra of nonthermal primary photon, and cutoff energies are lower [cf. Eq. (\ref{e_g0_max}); 
positions of cutoff are not seen in this Figure]. These results reflect the differential inverse Compton scattering rate [Eq. (\ref{eq_p_iC})].


\begin{figure}
\begin{center}
\includegraphics[width=8.0cm,clip]{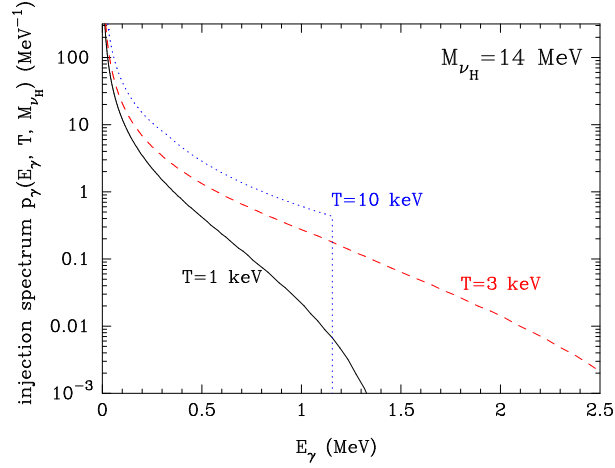}
\caption{Injection spectra of nonthermal photon formed through the electromagnetic cascade for $T=1$, 3 and 10 keV, respectively.  
The mass of the sterile neutrino is assumed to be $M_\nuh=14$ MeV. \label{fig6}}
\end{center}
\end{figure}

Figure \ref{fig6} shows injection spectra of nonthermal photon formed through the electromagnetic cascade calculated 
with Eq. (\ref{eq_inj}) for $T=1$, 3 and 10 keV, respectively.  As the temperature decreases, the primary photon spectrum $p_\gamma^{\rm pri}$ becomes softer so that less photons have high energies enough to disintegrate $^7$Be.
The spectrum of photons secondarily induced by the primary photons, however, has an upper cutoff at 
$E_C\sim m_e^2/(22T)$~\cite{Kawasaki:1994sc} which scales as inverse of the temperature.
Therefore, the cutoff energy is larger at lower temperature. The cutoff can be seen at 
$E_\gamma \sim 1.15$ MeV for $T=10$ keV.  Because of the combination of the softness of the primary photon spectrum and the cutoff energy in electromagnetic cascade shower, there is a best temperature of the energy injection 
where relatively large abundances of energetic photons are produced with $E_\gamma > 1.59$ MeV which can destroy $^7$Be.  
One can find that the best temperature is 3 keV among the three temperatures shown in this figure.


\subsection{light element abundances, $N_{\rm eff}$, and CMB energy spectrum}\label{sec6b}

\begin{figure}
\begin{center}
\includegraphics[width=8.0cm,clip]{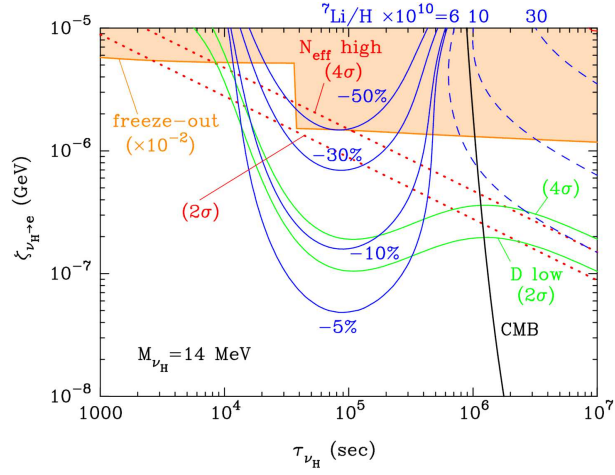}
\caption{Contours for calculated abundances of D and $^7$Li in the parameter plane of ($\tau_\nuh$, $\zeta_{\nuh\rightarrow e}$) 
for $M_\nuh=14$ MeV.
The regions above the curved diagonal solid lines marked as ``D low ($2\sigma$)'' and ``($4\sigma$)'' are excluded by D underproduction compared to the observational constraints at $2\sigma$ and $4\sigma$ levels, respectively.  The calculated $^7$Li abundance is smaller than the SBBN value above solid sharp curves at $\tau_\nuh \sim 10^5$ s by percentages shown near the curves, while it is larger above the dashed curves.  The dotted lines show the $2\sigma$ and $4\sigma$ upper limits on the effective neutrino number at the cosmological recombination epoch from CMB power spectrum.  The right region from the nearly-vertical solid line labeled as ``CMB'' is excluded from the limits from the CMB energy spectrum.  The shaded region corresponds to thermal freeze-out $\nuh$ abundances with a possible $\nuh$ dilution by a factor of 100 taken into account (See Sec. \ref{sec7a} and \ref{sec7b}).} \label{fig1}
\end{center}
\end{figure}

Figure \ref{fig1} shows contours for calculated abundances of D and $^7$Li in the ($\tau_\nuh$, $\zeta_{\nuh\rightarrow e}$) 
plane for $M_\nuh=14$ MeV.  The two curved diagonal solid lines marked as ``D low ($2\sigma$)'' and ``($4\sigma$)'' correspond to the observational $2\sigma$ and $4\sigma$ limits, respectively, on the D abundance.  
The regions above the lines are excluded by D underproduction.  Solid sharp curves at $\tau_\nuh \sim 10^5$ s are contours for the reduction ratio of $^7$Li abundance defined as
\begin{equation}
\Delta ^7{\rm Li} = \frac{\left(^7{\rm Li/H}\right) - \left(^7{\rm Li/H}\right)_{\rm SBBN}}{\left(^7{\rm Li/H}\right)_{\rm SBBN}},
 \label{eq_delta7li}
\end{equation}
where
($^7$Li/H) is the abundance calculated in this model, and ($^7$Li/H)$_{\rm SBBN}=5.07 \times 10^{-10}$ is the SBBN value.  Dashed curves are, on the other hand, contours for the calculated abundances ($^7$Li/H).  Inside the solid curves for $^7$Li, the calculated $^7$Li abundance is smaller than the SBBN value because of the $^7$Be photodisintegration.  The $^7$Li abundance is larger above the dashed curves than the SBBN value because of larger $\eta$ value during the BBN epoch.


The dotted lines show the $2\sigma$ and $4\sigma$ upper limits on the effective neutrino number at the cosmological recombination epoch from CMB power spectrum.  The nearly-vertical solid line labeled as ``CMB'' corresponds to the limits on the CMB energy spectrum.  
The right region from this line (long lifetime $\tau_\nuh \gtrsim 10^6$ s) is excluded from a large deformation in the energy spectrum.
We note that in the parameter region shown in Fig. \ref{fig1}, 
this model is constrained exclusively from the limit on the CMB $\mu$ parameter.

\begin{figure}
\begin{center}
\includegraphics[width=8.0cm,clip]{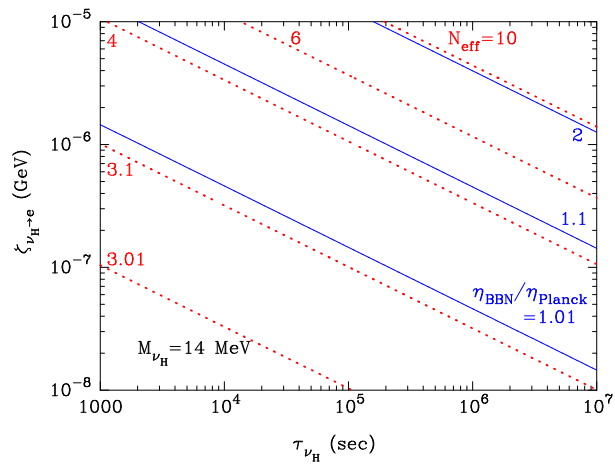}
\caption{Contours of the ratio between the baryon-to-photon ratios in the BBN epoch and the cosmological recombination epoch (solid lines), and contours of the $N_{\rm eff}$ value in the cosmological recombination epoch after the sterile neutrino decay in the parameter plane of ($\tau_\nuh$, $\zeta_{\nuh\rightarrow e}$) for $M_\nuh=14$ MeV.  \label{fig_neff}}
\end{center}
\end{figure}

Figure \ref{fig_neff} shows contours of the ratio between the baryon-to-photon ratios in the BBN epoch and the cosmological recombination epoch (solid lines), and contours of the $N_{\rm eff}$ value in the cosmological recombination epoch after the sterile neutrino decay in the parameter plane of ($\tau_\nuh$, $\zeta_{\nuh\rightarrow e}$) for $M_\nuh=14$ MeV.  


The parameter region for small $^7$Li abundance is found at $\tau_\nuh \sim 10^4-10^5$ s and $\zeta_{\nuh\rightarrow e}\sim 10^{-6}-10^{-7}$ GeV.  
In the epoch between the BBN and the matter radiation equality, the photon temperature in the universe $T$ is related to the cosmic age 
$t$ as $T=1.15$ keV $(t/10^6~{\rm s})^{-1/2}$ unless the cosmic expansion rate is significantly affected by the sterile neutrino.  If the $\nuh$ decay occurs at $T=\mathcal{O}$ (1 keV), 
the $^7$Li abundance is reduced most effectively although some amount of D destruction always occurs simultaneously.  
The shapes of contours of D and $^7$Li are explained as follows.  At short lifetimes of $\tau_\nuh \lesssim 10^4$ s, the upper cutoff of 
the nonthermal photon spectrum $E_C\sim m_e^2/(22T)$~\cite{Kawasaki:1994sc} is smaller than threshold energies for photodisintegration 
of light nuclides $E_{\gamma,{\rm th}} \sim {\cal O}(1-10)$ MeV.  Effects of nonthermal photons on elemental abundances are, therefore, negligibly small.  
At longer lifetimes of $\tau_\nuh \gtrsim 10^5$ s, on the other hand, primary photons produced via the inverse Compton scattering have softer 
energy spectra and  lower cutoff energy originating from the maximum energy of the scattered photon [Eq. (\ref{e_g0_max})].  
This fact results in smaller abundances of nonthermal photons which are energetic enough to destroy D and $^7$Li.  
As a result, effects of nuclear photodisintegration are less efficient in the long lifetime region also.

The lower solid line of D is located between the solid curves of $\Delta ^7$Li$=-5$ and $-10 \%$ in the interesting parameter region.  It means that in this region, the $^7$Be photodisintegration can slightly reduce 
the primordial $^7$Li abundance down to nearly the observational $2\sigma$ upper limit  multiplied by the stellar depletion factor of two.  
The D abundance, however, simultaneously decreases down to the observational $2\sigma$ lower limit.
We then find that there is no parameter region in which primordial $^7$Li abundance can be consistent with the MPS value without assuming a stellar depletion.  This model, however, provides a mechanism 
of $^7$Li reduction by some small factor with its signatures imprinted in the primordial D abundance and the effective neutrino number (Sec. \ref{sec6d}).

\begin{figure}
\begin{center}
\includegraphics[width=8.0cm,clip]{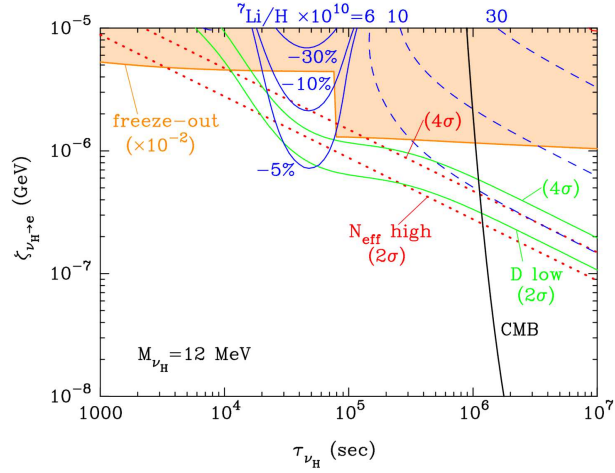}
\caption{Same as in Fig. \ref{fig1} but for $M_\nuh=12$ MeV. \label{fig2}}
\end{center}
\end{figure}

\begin{figure}
\begin{center}
\includegraphics[width=8.0cm,clip]{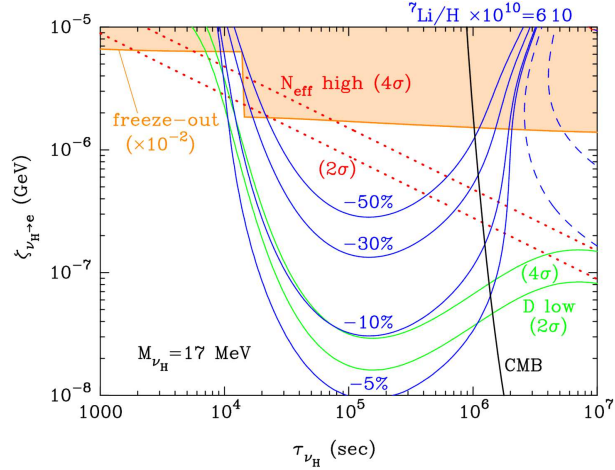}
\caption{Same as in Fig. \ref{fig1} but for $M_\nuh=17$ MeV. \label{fig3}}
\end{center}
\end{figure}

Figures \ref{fig2} and \ref{fig3} show contours for calculated abundances of D and $^7$Li, the effective neutrino number, and the CMB $\mu$ parameter as in Fig. \ref{fig1}, but for $M_\nuh=12$ and $17$ MeV, respectively.  
As seen in Figures \ref{fig1}, \ref{fig2}, and \ref{fig3}, regions of D and $^7$Li destruction at $\tau_\nuh \sim 10^5$ s are located at lower positions for larger masses.  This is because the sterile neutrinos with smaller masses can generate smaller numbers of energetic $e^\pm$' s which trigger D and $^7$Li destruction.  Since the energy fraction of energetic photons capable of destroying $^7$Be to total nonthermal photons is smaller, we need a larger total energy generated at the $\nuh$ decay.  Such a large energy injection is, however, constrained from BBN results taking into account the change of the baryon-to-photon ratio from the CMB-inferred value (Sec. \ref{sec4c}).  Therefore, for $M_\nuh < 14$ MeV the photodisintegration is less effective than the change of the baryon-to-photon ratio, and no parameter region for $^7$Li reduction is found.  The mass of $M_\nuh \gtrsim 14$ MeV is thus the best case for $^7$Be reduction that is a partial solution to the Li problem.  


The parameter region for the small primordial $^7$Li abundance is about to be excluded from observational constraint on D abundance.  In this parameter region, the change of D abundance partially originates from the change of the baryon-to-photon ratio.  In the case of smaller $M_\nuh$ values, the change of the baryon-to-photon ratio is more significant.  In the interesting parameter region, the effective neutrino number is also affected.  Therefore, this model for the $^7$Li reduction will be tested through observational determinations of the effective neutrino number (Sec. \ref{sec4b}) in near future.

\subsection{impact of revised cross sections}\label{sec6c}
We compare results of nonthermal nucleosynthesis calculated with the new and old cross sections of $^7$Be($\gamma$, $\alpha$)$^3$He and $^7$Li($\gamma$, $\alpha$)$^3$H.  Old cross sections have been adopted from the fitted functions in Ref. \cite{Cyburt:2002uv}.  For the reaction $^7$Be($\gamma$, $\alpha$)$^3$He, the cross section is set to be zero for $5.41$ MeV$<E_\gamma$ since the function gives negative values.

\begin{figure}
\begin{center}
\includegraphics[width=8.0cm,clip]{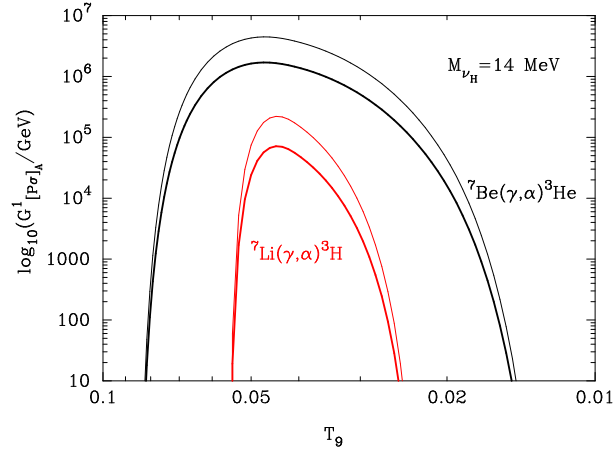}
\caption{Transfer functions $G_{[P\gamma]_A}^1(T; 14~{\rm MeV})$ [Eq. (\ref{nonthermal_rate1b})] calculated with the new (thick lines) and old (thin lines) cross sections of $^7$Be($\gamma$, $\alpha$)$^3$He and $^7$Li($\gamma$, $\alpha$)$^3$H as a function of $T_9$. \label{g_func}}
\end{center}
\end{figure}

Figure \ref{g_func} shows the transfer functions $G_{[P\gamma]_A}^1(T; 14~{\rm MeV})$ [Eq. (\ref{nonthermal_rate1b})] calculated with the new (thick lines) and old (thin lines) cross sections as a function of $T_9$.  The new transfer function of $^7$Be($\gamma$, $\alpha$)$^3$He is 2.3 times smaller than the old one, while that of $^7$Li($\gamma$, $\alpha$)$^3$H is 2.5 times smaller than the old one at their peak positions.  These changes are caused mainly by the fact that the new cross sections at low energies near the photodisintegration threshold energies are about one third of the old ones (Fig. \ref{fig_a1}).


\begin{figure}
\begin{center}
\includegraphics[width=8.0cm,clip]{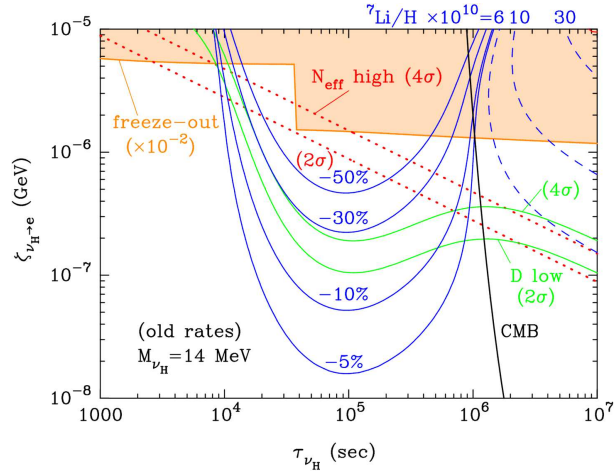}
\caption{Same as in Fig. \ref{fig1} for $M_\nuh=14$ MeV, but the old cross sections of $^7$Be($\gamma$, $\alpha$)$^3$He and $^7$Li($\gamma$, $\alpha$)$^3$H \cite{Cyburt:2002uv} are used in the calculation. \label{fig1_old}}
\end{center}
\end{figure}

Figure \ref{fig1_old} shows the same contours for calculated abundances of D and $^7$Li, the effective neutrino number, and the CMB $\mu$ parameter as in Fig. \ref{fig1} for $M_\nuh=14$ MeV, but for results calculated with old cross sections.  Solid lines for $^7$Li are located in the lower positions than those in Fig. \ref{fig1}.  This is because the adopted cross sections of $^7$Be($\gamma$, $\alpha$)$^3$He and $^7$Li($\gamma$, $\alpha$)$^3$H are smaller, and resultingly, the photodisintegration rates are smaller (Fig. \ref{g_func}).  We note that when old cross sections are used carelessly, one finds a fake parameter region in which $^7$Li abundance can be significantly reduced without a large efficiency of D photodisintegration, as shown in this figure.  It is, therefore, necessary to adopt the precise cross sections of the reactions $^7$Be($\gamma$, $\alpha$)$^3$He and $^7$Li($\gamma$, $\alpha$)$^3$H.


\begin{figure}
\begin{center}
\includegraphics[width=8.0cm,clip]{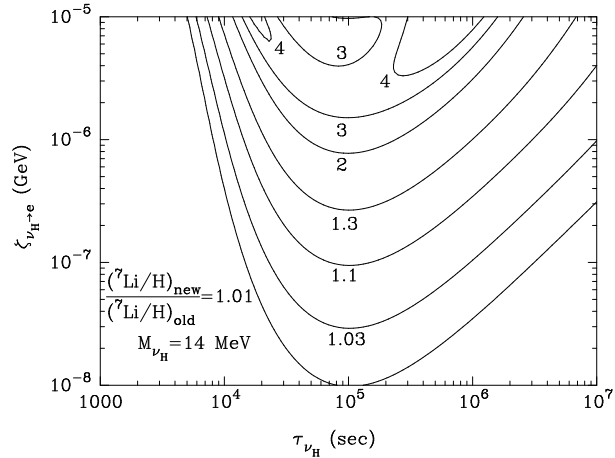}
\caption{Ratio between the $^7$Li/H abundances calculated with the new and old cross sections of $^7$Be($\gamma$, $\alpha$)$^3$He and $^7$Li($\gamma$, $\alpha$)$^3$H in the parameter plane of ($\tau_\nuh$, $\zeta_{\nuh\rightarrow e}$) for $M_\nuh=14$ MeV.  \label{fig_li7}}
\end{center}
\end{figure}

Figure \ref{fig_li7} shows contours of the ratio between the $^7$Li/H abundances calculated with the new and old cross sections of $^7$Be($\gamma$, $\alpha$)$^3$He and $^7$Li($\gamma$, $\alpha$)$^3$H in the parameter plane of ($\tau_\nuh$, $\zeta_{\nuh\rightarrow e}$) for $M_\nuh=14$ MeV.  In the parameter region at $\tau_\nu \sim 10^5$ s, the photodisintegration of $^7$Be and $^7$Li is efficient.  Since the photodisintegration rates from the new cross sections are smaller than those from the old ones, larger abundances of $^7$Be and $^7$Li survive the photodisintegration.  The final abundance of $^7$Li, given by the sum of the $^7$Be and $^7$Li abundances, is, therefore, larger in the case of the new cross sections.


\subsection{parameter region for $^7$Be destruction}\label{sec6d}
We take a parameter set for the $^7$Be destruction: ($\tau_\nuh$, $\zeta_{\nuh\rightarrow e}$)=($4\times 10^4$ s, $3\times 10^{-7}$ GeV) in the case of $M_\nuh=14$ MeV.

\begin{figure}
\begin{center}
\includegraphics[width=8.0cm,clip]{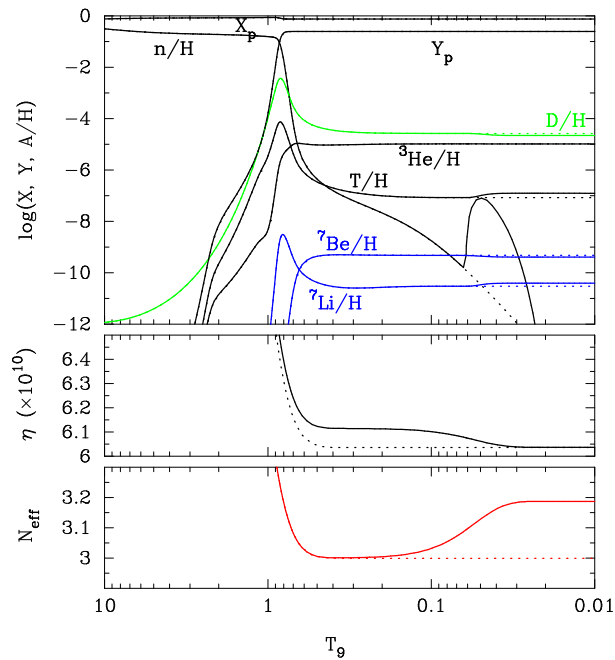}
\caption{Nuclear abundances (the top panel), the baryon-to-photon ratio (the middle panel), and the effective neutrino number [Eq. (\ref{eq_neff})] (the bottom panel) as a function of $T_9$.  The solid and dotted lines show results of this model with the decaying sterile neutrino and the SBBN, respectively.  In the top panel, $X_{\rm p}$ and $Y_{\rm p}$ are the mass fractions of $^1$H and $^4$He, respectively, while abundances of other nuclides are given by ratios of number densities of nuclides and hydrogen.  The mass, lifetime, and abundance of the sterile neutrino are set to be $M_\nuh=14$ MeV, $\tau_\nuh=4 \times 10^4$ s, and $\zeta_{\nuh \rightarrow e}=3\times 10^{-7}$ GeV, respectively.  \label{fig_evolution}}
\end{center}
\end{figure}

Figure \ref{fig_evolution} shows calculated nuclear abundances (the top panel), the baryon-to-photon ratio (the middle panel), and the effective neutrino number [Eq. (\ref{eq_neff})] (the bottom panel) as a function of $T_9$.  The solid and dotted lines correspond to results of the present model with the decaying sterile neutrino and the SBBN, respectively.  In the top panel, $X_{\rm p}$ and $Y_{\rm p}$ are the mass fractions of $^1$H and $^4$He, respectively, while other curves are number densities of other nuclides relative to that of hydrogen.


In the top panel, effects of photodisintegration can be observed as differences of solid and dotted lines at $T_9 \lesssim 0.06$.  The $^7$Be nuclei are disintegrated via $^7$Be($\gamma$, $\alpha$)$^3$He, and the $^7$Be abundance is slightly decreases.  The deuterons are also disintegrated via $^2$H($\gamma$, $n$)$^1$H, and the D abundance decreases and the neutron abundance increases.  When the photodisintegration occurs, the temperature is already enough low that thermal nuclear reactions between charged nuclei are no longer operative.  However, nonradiative neutron capture reactions are operative since no Coulomb repulsion exists in reactions with neutrons.  The generated neutrons are, therefore, partially captured by $^3$He [via $^3$He($n$, $p$)$^3$H] and $^7$Be [via $^3$Be($n$, $p$)$^7$Li].  As a result, abundances of $^3$H and $^7$Li slightly increase.

In the middle panel, the baryon-to-photon ratio $\eta$ decreases at $T_9 \lesssim 0.1$ because of an electromagnetic energy injection at the $\nuh$ decay.  Since the final $\eta$ value is fixed to the Planck value, the baryon-to-photon ratio in the $\nuh$ model is higher than that in SBBN at $T_9\gtrsim 0.1$.  Although this difference of the $\eta$ value slightly changes BBN, differences in nuclear abundances during BBN epoch ($T_9 \lesssim 1$) are so small that they cannot be seen well.  The downturns of the solid and dotted lines at $T_9 \gtrsim 1$ are caused by an entropy transfer from $e^\pm$ to photon at the cosmological $e^\pm$ annihilation.  The $\eta$ values before the $e^\pm$ annihilation are $1.68\times 10^{-9}$ in the $\nuh$ model and $1.66\times 10^{-9}$ in the SBBN model, respectively.

In the bottom panel, the solid and dashed lines descend at $T_9 \gtrsim 1$ by the entropy transfer from $e^\pm$ to photon.  The $N_{\rm eff}$ value is then increased at $T_9 \lesssim 0.1$ by the nonthermal neutrino injection at the $\nuh$ decay.  The final $N_{\rm eff}$ values after the $\nuh$ decay are 3.19 in the $\nuh$ model and 3 in the SBBN model, respectively.

We check the validity of the approximate formula for the final $N_{\rm eff}$ value.
The present lifetime corresponds to $t_{\rm eff}=(\pi/4) \tau_\nuh =3.14\times 10^4$ s 
(Sec. \ref{sec4d2}) and $T(t_{\rm eff})=6.48$ keV.
The present mass corresponds to parameters of $x_{\rm m}=0.0365$ and $L=-13.236$ [Eq. (\ref{eq_a12})].  
The ratio of decay rates is $R_{\rm dec}=0.575$ [Eq. (\ref{eq_R_dec})].
The ratios of the average energies to the sterile neutrino mass in the decay mode $\nuh\rightarrow \nu_e e^+ e^-$ are 
$f_E(\nu_e)=0.348$, $f_E(e^-)=0.308$, and $f_E(e^+)=0.345$.
The ratio of injected energies of neutrinos and $e^\pm$ is then derived as $R(\nu,e)=3.20$ [Eq. (\ref{eq_ratio_nue})].
The corresponding ratio of entropy densities before and after the $\nuh$ decay is $S_{\gamma,{\rm aft}}/S_{\gamma,{\rm bef}}=1.0129$ [Eq. (\ref{eq_entropy})].
The effective neutrino number contributed from nonthermal neutrinos without the entropy production effect is derived as $N_{\rm eff,nt,i}=0.241$ [Eqs. (\ref{eq_zeta_ratio}) and (\ref{Neff_nt_i})].
Taking account of the dilution associated with the entropy production, the final effective number is $N_{\rm eff}=3.19$ [Eq. (\ref{eq_dn})].
We note that a slight change in the cosmic expansion rate through a change in the $g_\ast$ value by the $\nuh$ was neglected in this estimation.  It is thus found that this approximate estimation successfully gives the final $N_{\rm eff}$ value derived in our precise numerical calculation.

\section{Discussions}\label{sec7}
\subsection{relic abundance of sterile neutrino}\label{sec7a}
It is assumed that the sterile neutrino has a mass $M_\nuh$ after the EW phase transition.  Depending on the mixing angle, the sterile neutrino react with standard model particles mainly via the following weak reactions:
\begin{eqnarray}
\nuh + \nu_e &\rightarrow& f +\bar{f} \label{process_1}\\ 
\nuh + f &\rightarrow& \nu_e +f \\
\nuh + \bar{f} &\rightarrow& \nu_e +\bar{f} \\
\nuh + e^+ &\rightarrow& \bar{e_n}^+ + \nu_{e_n} \\
\nuh + e^+ &\rightarrow& u_n^{+2/3} + \bar{d_{n^\prime}}^{+1/3} \\
\nuh + e_n^- &\rightarrow& e^- + \nu_{e_n} \\
\nuh + \bar{\nu}_{e_n} &\rightarrow& e^- + \bar{e_n}^+ \\
\nuh + \bar{u_n}^{-2/3} &\rightarrow& e^- + \bar{d_{n^\prime}}^{+1/3} \\
\nuh + d_n^{-1/3} &\rightarrow& e^- + u_{n^\prime}^{+2/3}, \label{process_9}
\end{eqnarray}
where
$f$ is any fermion, i.e., charged leptons $e_n$ [$e^-$ ($n=1$), $\mu^-$ ($n=2$), and $\tau^-$ ($n=3$)], neutrinos $\nu_{e_n}$ and up-type quarks $u_n$ [$u$ ($n=1$), $c$ ($n=2$), and $t$ ($n=3$)], and down-type quarks $d_n$ [$d$ ($n=1$), $s$ ($n=2$), and $b$ ($n=3$)].  In the charged current reactions, probabilities of producing respective flavors ($n_n \leftrightarrow d_{n^\prime}$) are described by the Cabibbo-Kobayashi-Maskawa matrix \cite{Cabibbo:1963yz,Kobayashi:1973fv}.

When the weak interaction rate becomes smaller than the Hubble expansion rate, the abundance of the sterile neutrino freezes out from equilibrium.  Thereafter, the ratio between the $\nuh$ number density and the entropy density $Y_\nuh\equiv n_\nuh/s$ does not change \footnote{We note that this ratio $Y_\nuh$ is measured in a unit different from that of nuclear mole fractions $Y_A$ introduced in Sec. \ref{sec2g}.}.
Weak interaction rates of sterile neutrinos, $\nuh$'s, with weakly interacting standard model particles 
after the EW phase transition scale \cite{Dolgov:2000pj} as
\begin{equation}
\Gamma \sim G_{\rm F}^2 \Theta^2 T^5,
\end{equation}
where
$G_{\rm F}$ is the Fermi constant,
$\Theta\ll 1$ is the mixing angle.
The ratio between the rates and cosmic expansion rate, $H$, is then given \cite{Fuller:2011qy} by
\begin{eqnarray}
\frac{\Gamma}{H} &\sim& G_{\rm F}^2 \Theta^2 T^5 \left( \frac{2\pi^{3/2}}{3 \sqrt[]{\mathstrut 5}} 
\frac{g_\ast^{1/2} T^2}{M_{\rm Pl}} \right)^{-1} \nonumber\\
&=& \frac{3 \sqrt[]{\mathstrut 5}}{2\pi^{3/2}} \frac{M_{\rm Pl}G_{\rm F}^2 
\Theta^2 T^3}{g_\ast^{1/2}} \nonumber\\
&=& 9.69 \times 10^{7} \left(\frac{\Theta}{10^{-3}} \right)^2 \left(\frac{g_\ast}{106.75}\right)^{-1/2} 
\left(\frac{T}{100~{\rm GeV}}\right)^3 \nonumber\\
&=& 1.00 \left(\frac{\Theta}{10^{-3}} \right)^2 \left(\frac{g_\ast} {63.75}\right)^{-1/2} 
\left(\frac{T}{0.2~{\rm GeV}}\right)^3. \label{eq_temp_freezeout}
\end{eqnarray}
In the last line of this equation, the statistical DOF of the sterile neutrino, i.e., $\Delta g_\ast=2\times 7/8$ 
were added to the value of $g_\ast=61.75$ at $T=200$ MeV in the standard model.
A sterile neutrino with mixing angle $\Theta\sim 10^{-3}$ would thus freeze out from equilibrium 
at temperature $T\sim 200$ MeV.  The relic abundance of $\nuh$ is, therefore, given by the abundance 
fixed at $T\sim 200$ MeV.

The lifetime of the sterile neutrinos is roughly given [cf. Eqs. (\ref{eq_a9}) and (\ref{eq_a21})] by 
\begin{eqnarray}
\Gamma(\nuh-{\rm decay}) &\sim& \frac{G_{\rm F}^2 \Theta^2 M_\nuh^5}{192 \pi^3} \nonumber\\
&=& 1.87 \times 10^{-5}~{\rm s}^{-1} \left( \frac{\Theta}{10^{-3}}\right)^2 \left( \frac{M_\nuh} {14~{\rm MeV}}\right)^5.
\label{gamma_approx}
\end{eqnarray}
Therefore, if a sterile neutrino with a mass $M_\nuh \gtrsim 14$ MeV decays $\sim 10^4 -10^5$ s after the big bang and reduces the primordial $^7$Li abundance, the mixing angle would be $\Theta \sim 10^{-3}$.

The time evolution of the $\nuh$ abundance has been calculated \cite{Kolb:1981cx,Scherrer:1987rr,Dolgov:2000pj}.  In Refs. \cite{Kolb:1981cx,Scherrer:1987rr}, however, the maximal mixing angle $\Theta ={\mathcal O}(1)$ is implicitly assumed, and the dependence on the mixing angle $\Theta$ is not considered.  In addition, the authors took into account only the annihilation $\nuh +\bar{\nuh}$ \cite{Dicus:1977qy}, which is negligibly weaker than the reactions Eqs. (\ref{process_1})-(\ref{process_9}) when $\Theta \ll 1$.  In Ref. \cite{Dolgov:2000pj}, on the other hand, a dedicated calculation has been performed.  However, the authors only focused on shorter $\nuh$ lifetimes of $\tau_\nuh={\mathcal O}(0.1)$ s, which correspond to relatively large values of the mixing angle $\Theta >{\mathcal O}(10^{-3})$ compared with those considered in this paper.  Depending on the mixing angle, the weak reaction freeze-out of the sterile neutrino occurs in various epochs with different values of $g_\ast$.  Perhaps the sterile neutrino never experiences the weak reaction equilibrium after the EW phase transition.  In general, the $\nuh$ relic abundance can sensitively depend on the evolution of sterile neutrino mass during the EW phase transition, which differs in different models of the sterile neutrino.  Precise calculations of the $\nuh$ relic abundance should be performed in detail.  However, they are beyond the scope of this paper.

We estimate the freeze-out abundance of the sterile neutrino as a function of $M_\nuh$ and $\tau_\nuh$ as follows.  For a given set of ($M_\nuh$, $\tau_\nuh$), a corresponding $\Theta$ value is derived with Eq. (\ref{gamma_approx}).  The temperature satisfying Eq. (\ref{eq_temp_freezeout}) is then derived with the $\Theta$ value.  This temperature is defined as the freeze-out temperature $T_{\rm F}$.  An approximate value of the freeze-out abundance of $\nuh$ is given by the equilibrium abundance at $T_{\rm F}$.   For the $T_{\rm F}(M_\nuh, \tau_\nuh)$ value, the freeze-out abundance is given by the equilibrium abundance $Y_{\nuh,{\rm EQ}}(M_\nuh, T_{\rm F})$ using the following equations. 

The equilibrium number density of a fermion is given \cite{kolb1990} by
\begin{equation}
n_{i,{\rm EQ}}(m_i,T)= \frac{g_iT^3}{2\pi^2} h(m_i/T),
\label{n_eq}
\end{equation}
where
$m_i$ and $g_i$ are the mass and statistical DOF, respectively, of the fermion $i$, and $h(x)$ is a function given by
\begin{equation}
h(x)= \int_x^\infty \frac{\left( \epsilon^2 -x^2 \right)^{1/2} \epsilon}{\exp\left( \epsilon \right) +1} d\epsilon.
\label{h_plus}
\end{equation}
In the nonrelativistic limit, the function $h$ has the limit value of $h \rightarrow 3\zeta(3)/2$.  The entropy density is given \cite{kolb1990} by
\begin{equation}
s(T)= \frac{2 \pi^2}{45} g_{\ast{\rm S}} T^3.
\label{eq_entropy_density}
\end{equation}
From Eqs. (\ref{n_eq}) and (\ref{h_plus}), the abundance ratio is given by
\begin{equation}
Y_{i,{\rm EQ}}(m_i,T) \equiv \frac{n_{i,{\rm EQ}}(m_i,T)}{s(T)}= \frac{45 g_i}{4\pi^4 g_{\ast{\rm S}}} h(m_i/T).
\label{eq_entropy_density}
\end{equation}

\begin{figure}
\begin{center}
\includegraphics[width=8.0cm,clip]{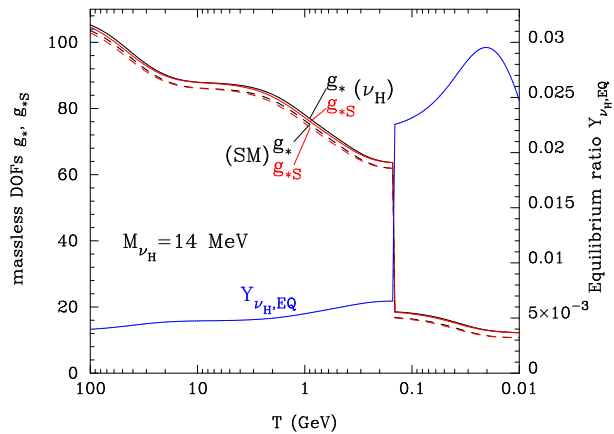}
\caption{Massless DOFs in terms of energy and entropy, i.e., $g_\ast$ and $g_{\ast{\rm S}}$, respectively, as a function of photon temperature $T$.  Solid lines for massless DOFs correspond to the standard model plus a sterile neutrino of mass $M_\nuh=14$ MeV and statistical DOF of $g_\nuh=2$, while dashed lines correspond to the standard model.  Also shown is the equilibrium abundance ratio of a sterile neutrino $Y_{\nuh,{\rm EQ}}$ in the model with the sterile neutrino.  \label{fig_relic}}
\end{center}
\end{figure}

Figure \ref{fig_relic} shows massless DOFs in terms of energy and entropy, i.e., $g_\ast$ and $g_{\ast{\rm S}}$, respectively, as a function of photon temperature $T$.  Solid lines for massless DOFs correspond to the standard model plus a sterile neutrino of mass $M_\nuh=14$ MeV and statistical DOF of $g_\nuh=2$, while dashed lines correspond to the standard model.  Also shown is the equilibrium abundance ratio of a sterile neutrino $Y_{\nuh,{\rm EQ}}$ in the model with the sterile neutrino.  The massless DOFs are calculated as in Ref. \cite{kolb1990} based on the latest data on particle mass \cite{Beringer:1900zz}.  It is assumed that the quark hadron transition occurs suddenly at temperature $T_{\rm C}=150$ MeV.  Above the temperature, quarks are taken into account in the DOFs.  Below the temperature, on the other hand, contributions of only hadrons are included and those of quarks are neglected.  We only take into account DOFs of charged and neutral pions at $T<T_{\rm C}$ since they are only relativistic hadrons.


The massless DOFs in the model with $\nuh$ is larger than those in the model without $\nuh$ by about two because of the statistical DOF of sterile neutrino.  As the temperature decreases, weak bosons, heavy quarks and leptons become nonrelativistic, and the DOFs become small.  At the quark hadron transition temperature $T=T_{\rm C}$, DOFs of quarks and gluons disappear, and the DOFs drastically decreases.  The equilibrium abundance $Y_{\nuh,{\rm EQ}}$ increases as the temperature increases since it is proportional to the $g_{\ast{\rm S}}(T)$ value.  At $T_{\rm C}$, the abundance significantly increases.  At the lowest temperature of $T\lesssim 20$ MeV, the sterile neutrino start to be nonrelativistic.  The equilibrium abundance then decreases from this temperature.

If a light sterile neutrino with the mass $M_\nuh ={\cal O}(10)$ MeV survives during BBN epoch, 
its number density must have diluted between its weak freeze-out [$T\sim {\cal O}(100)$ MeV] and the BBN epoch [$T\sim {\cal O}(0.1)$ MeV] in order to avoid a large change of the baryon-to-photon ratio associated with the $\nuh$ decay (see Sec. \ref{sec7b}).  For example, we consider the case of $M_\nuh=14$ MeV and 
$\zeta_{\nuh\rightarrow e}=3\times 10^{-7}$ GeV.
This assumption corresponds to the energy ratio $\zeta_{\nuh\rightarrow e}/\zeta_{\nuh\rightarrow \nu}=0.313$ [Eqs. (\ref{eq_ratio_nue}) and (\ref{eq_zeta_ratio})], and 
the total energy injection of 
$\zeta_{\nuh}=\zeta_{\nuh\rightarrow e}+\zeta_{\nuh\rightarrow \nu}=1.26\times 10^{-6}$ GeV.
This energy injection is realized by the decay of sterile neutrino with the mass $M_\nuh=14$ MeV and 
the number ratio $Y_\nuh=1.28\times 10^{-5}$, where we used a relation,
\begin{equation}
\zeta_{\nuh} =\frac{n_\nuh}{s} \frac{s}{n_\gamma} M_\nuh =7.04Y_\nuh M_\nuh,
\label{relation_zeta_Y}
\end{equation}
where the ratio $s/n_\gamma=7.04$ should be measured after the cosmological $e^\pm$ annihilation.
However, the freeze-out abundance is 
$Y_\nuh=6.56\times 10^{-3} (g_{\ast{\rm S}}/63.5)^{-1}$.
Therefore, the sterile neutrino needs to be diluted by a factor of several hundreds.

\subsection{dilution of sterile neutrino}\label{sec7b}
It is shown that the decays of heavier sterile neutrinos into standard model particles can realize a dilution of the light sterile neutrino 
($M_\nuh\sim 14$ MeV) although our naive estimate~\cite{Asaka:2006nq} suggests that a decay of a heavier sterile neutrino would 
result in a dilution factor smaller than required for the appropriate abundance 
by some factor at least.
We assume that one of heavier sterile neutrinos, i.e., $\nuh_2$, predominantly contributes to the dilution or entropy production.  In addition, it is assumed that the heavy neutrino dominates in terms of energy density in its decay epoch.
Supposing that $\nuh_2$ decays into relativistic leptons and quarks which are thermalized rapidly with respect 
to the cosmic expansion time scale, the energy density of relativistic species after the decay is
$\rho_{\rm R} = \frac{\pi^2}{30} g_\ast T_{\rm RH}^4$,
where
$T_{\rm RH}$ is the reheating temperature.
This energy density is equal to the energy density of $\nuh_2$ before the decay.  
The ratio of the entropy per comoving volume at the epoch long after the decay (aft) to that long before the decay 
(bef) is given \cite{kolb1990} by 
\begin{eqnarray}
\frac{S_{\rm aft}}{S_{\rm bef}} &=& \frac{g_{\ast{\rm S}}(T_{\rm aft}) a_{\rm aft}^3 T_{\rm aft}^3}{g_{\ast{\rm S}}(T_{\rm bef}) a_{\rm bef}^3 T_{\rm bef}^3} \nonumber\\
&\simeq & 1.83\langle g_\ast^{1/3} \rangle^{3/4} \frac{m_{\nuh_2} Y_{\nuh_2} \tau_{\nuh_2}^{1/2}}{M_{\rm Pl}^{1/2}} \nonumber\\
&= & 8.25 \times 10^1 \left( \frac{\langle g_\ast^{1/3} \rangle^{3/4}}{104^{1/4}} \right) \left( \frac{m_{\nuh_2}}{100~{\rm GeV}} \right) \left( \frac{Y_{\nuh_2}}{4.00\times 10^{-3}}
\right) \left( \frac{\tau_{\nuh_2}}{10^{-2}~{\rm s}} \right)^{1/2},
\end{eqnarray}
where 
$a_j$ and $T_j$ (for $j=$bef and aft) are the scale factor and the photon temperature of the universe at time $j$, and 
we supposed $g_\ast=g_{\ast{\rm S}}$ and that the $g_\ast$ and $g_{\ast{\rm S}}$ values do not change 
between the temperatures of $T_{\rm bef}$ and $T_{\rm aft}$.  
We note that a large dilution factor is realized only for $M_{\nuh_2}\lesssim 100$ GeV.  If the mass is much larger than the energy scale of the EW phase transition, the freeze-out $\nuh_2$ abundance is small because of the Boltzmann suppression factor.  Furthermore, the lifetime should not be longer than ${\mathcal O}(10^{-2})$ s since BBN is significantly affected by nonthermal reactions of hadronic particles generated at the $\nuh_2$ decay if the lifetime is longer \cite{Kawasaki:2004qu}.
From this equation, we find that the dilution factor, 
that equals the entropy enhancement factor, is about a factor of 100 at maximum.  This maximum factor is $\sim$3 times smaller than the necessary factor of 300.  Some other mechanism of the dilution is, therefore, 
needed for the light sterile neutrino to destroy some moderate fraction of primordial $^7$Be successfully.
Possible mechanisms include dilutions by massive particles other than $\nuh_2$ decaying into active neutrinos, $\phi \rightarrow \nu \bar{\nu}$ \cite{Hannestad:2004px} or photons $\phi \rightarrow n\gamma$ ($n \geq 2$) \cite{Ichikawa:2005vw}.

In Figs. \ref{fig1}, \ref{fig2}, \ref{fig3}, and \ref{fig1_old}, parameter regions of thermal freeze-out $\nuh$ abundances are shown by shaded regions.  A possible dilution of the sterile neutrino by a factor of 100 is taken into account.  The freeze-out abundances without dilutions are higher than the figure domains, and therefore not seen.  The lower boundaries of these regions correspond to the abundances diluted by a factor of 100.  Sudden drops of the boundaries at $\tau_\nuh =10^4-10^5$ s result from the decrease in the massless DOF in terms of entropy (Fig. \ref{fig_relic}).  It is clear that the parameter regions for the primordial $^7$Li reduction are lower than the regions of freeze-out abundances.  Therefore, a dilution of the sterile neutrino is necessary for the $^7$Li reduction to work.

\subsection{mixing with muon and tauon neutrinos}\label{sec7c}
If either muon or tauon neutrino predominantly couples to $\nuh$ and couplings of other charged leptons are negligible as an extreme case opposite to the case studied in this paper, effects of the sterile neutrino decay are changed.  For example, we take the case of $M_\nuh=14$ MeV.  The ratio of energy injections in the forms of $e^\pm$ and $\nu$ is $R(\nu, e)=3.20$ for the coupling to $\nu_e$, while it is $R(\nu, \alpha)=13.0$ for the coupling to $\nu_\alpha$ ($\alpha=\mu$ or $\tau$) [Eq. (\ref{eq_zeta_ratio})].  The muon or tauon type mixings, therefore, result in a large energy fraction of neutrino emitted at the decay.  An $e^\pm$ injection decreases the $\eta$ and $N_{\rm eff}$ values and induces nonthermal nucleosynthesis, while a neutrino injection increases the $N_{\rm eff}$ value.  Therefore, a sterile neutrino that mixes only with $\nu_\alpha$ has small effects on the primordial light element abundances and the $\eta$ value relative to that on $N_{\rm eff}$.  

\subsection{constraint from pion decay}\label{sec7d}
We assume that the sterile neutrino has a mass $\sim 14$ MeV and a lifetime $\sim 10^5$ s (parameter value for the $^7$Li reduction), and that mixing angles of muon and tauon types, $\Theta_\mu$ and $\Theta_\tau$, respectively, can be neglected. 
The active-sterile mixing angle is then determined to be $\Theta ={\cal O}(10^{-3})$ [Eqs. (\ref{gamma_approx}) and (\ref{eq_a9})].  
If the muon type mixing is sizable, we should take into account another
constraint from low energy phenomena: the sterile neutrino can be
produced by the decay of charged pions,
e.g., $\pi^+ \to \mu^+ + \nuh$ or $\pi^+ \to e^+ + \nuh$.
This channel has been searched for a long time and many experiments
gave constraints on the active-sterile mixing angle.
From Ref.~\cite{Kusenko:2004qc}, $\Theta^2_\mu$ should be smaller than $10^{-5}$.
If the precision of those experiments can be improved by a few orders
of magnitude, therefore,
we may see a signal from pion decays, or
exclude the possibility of primordial $^7$Li reduction by $\nuh$ suggested in this paper.
Furthermore,
if the muon type mixing is of the order of $\Theta_\mu \sim 10^{-6}$,
the sterile neutrino might be detected by
Super Kamiokande in future~\cite{Asaka:2012hc}.
It is worth mentioning
that a constraint from supernova SN1987A  observation
is rather strong~\cite{Dolgov:2000pj,Dolgov:2000jw};
that is roughly $\Theta^2 \lesssim {\cal O}(10^{-8})$ for any flavors. 

\section{Summary}\label{sec8}

The primordial lithium abundance determined from spectroscopic observations of MPSs is smaller than the 
theoretical prediction of SBBN model by a factor of $\sim 3$. It has been suggested that a BBN model with a long-lived radiatively 
decaying exotic particle possibly provides a solution to the Li problem.  In that model nonthermal photons with energies $\sim 2$ MeV generated 
by the particle decay disintegrate $^7$Be.  The primordial $^7$Li abundance, which is the sum of abundances 
of $^7$Li and $^7$Be produced during BBN, is then reduced.  In this paper, we studied 
the possibility of $\mathcal{O}(10)$ MeV sterile neutrino $\nuh$.  If it decays after BBN, and electron and positron 
$e^\pm$'s are emitted, the energetic $e^\pm$'s can produce energetic photons via the inverse Compton scattering 
of CBR.  The solution to the Li problem is, therefore, also expected in this model.  Then, we investigated cosmological effects of the sterile neutrino decay on primordial light element abundances, the baryon-to-photon ratio, and the effective neutrino number.

The sterile neutrino was assumed to live longer than the BBN time scale, i.e., $\gg \mathcal{O}(100)$ s.  This long lifetime satisfies a recent constraint from neutrino oscillation experiments:  the lifetime of sterile neutrino lighter than the pion cannot be shorter than $\sim 0.1$ s.  We constructed a numerical code for consistent calculations of the cosmic expansion history, BBN, and nonthermal nucleosynthesis triggered by the $\nuh$ decaying after BBN.  The updated relation between the baryon-to-photon ratio and the baryon density of the universe is used (Sec. \ref{sec2a1}).  The initial abundance, mass, and lifetime of the sterile neutrino were taken as free parameters.  Then we formulated the injection spectrum of nonthermal photon induced by the $\nuh$ decay (Sec. \ref{sec2}).  We introduced an active-sterile mixing angle, and calculated the energy spectra of $e^\pm$'s and active neutrinos generated at the $\nuh$ decay.  Taking into account the primary photon production via the inverse Compton scattering of CBR by energetic $e^\pm$'s, and electromagnetic cascade showers induced by the primary photons, the steady state injection spectrum was derived as a function of the sterile neutrino mass and the photon temperature.  Nonthermal nucleosynthesis triggered by the energetic photons are then calculated.  In this paper, we corrected errors in photodisintegration cross sections of $^7$Be($\gamma$, $\alpha$)$^3$He and $^7$Li($\gamma$, $\alpha$)$^3$H adopted in previous studies.  We gave functions for the cross sections in light of recent nuclear experimental results (Sec. \ref{sec3}).  Furthermore, effects of the $\nuh$ decay on the cosmic thermal history and evolutions of effective neutrino number and the baryon-to-photon ratio are formulated exactly (Sec. \ref{sec4}).  

Results of our calculations are summarized as follows:

First, we calculated 
injection spectra of nonthermal photon as a function of the mass $M_\nuh$ and 
photon temperature $T$.  We took into account the energy spectra of $e^\pm$'s emitted at the decay, inverse Compton 
scattering of CBR by the energetic $e^\pm$'s producing primary energetic photons, and electromagnetic cascade 
showers induced by the primary photons.  The energy spectra of $e^\pm$'s are broadly extended independent of 
the temperature.  The energy spectra of the primary photons, on the other hand, depend significantly on the temperature, 
and spectra are softer at lower temperatures in the later Universe.  The final injection spectra of nonthermal photon also 
depend on the temperature significantly.  Abundances of energetic photons capable of disintegrating
  $^7$Be are determined by the hardness of the primary photon spectra and the upper cutoff in the nonthermal photon spectra 
due to the double photon pair annihilation.  We found that an effective $^7$Be destruction can occur only if the sterile neutrino decays at $T={\cal O}(1)$ keV  (Sec. \ref{sec6a}).

Second, we simultaneously solved nonthermal nucleosynthesis induced by the nonthermal photons, and evolutions of the baryon-to-photon ratio $\eta$ and the cosmological effective neutrino number $N_{\rm eff}$.  
At the $\nuh$ decay, energetic active neutrinos, electrons, and positrons are generated.  
The energies of the neutrinos are never thermalized since the weak interaction has been long since decoupled in the universe.  The nonthermal neutrinos, therefore, contribute to only the radiation energy density or the $N_{\rm eff}$ value.  The energies of the $e^\pm$'s are, on the other hand, quickly thermalized through interactions with CBR, and eventually transferred to CBR.  The comoving photon entropy is then increased.  
Using formulae relevant to the sterile neutrino decay (Appendix \ref{app1}), we quantitatively solved changes of $\eta$ and $N_{\rm eff}$ caused by the $\nuh$ decay.  The final $\eta$ value is fixed to the Planck value at the cosmological recombination.  Calculated results are compared with observational constraints.  As a result, amounts of energy injection in the form of $e^\pm$'s at the $\nuh$ 
decay are constrained from limits on primordial nuclear abundances (D and $^7$Li), the effective neutrino number, and the CMB energy spectrum.  We found a parameter region of the lifetime $\tau_\nuh$ and the amount of energy injection 
$\zeta_{\nuh\rightarrow e}$, in which $^7$Be is photodisintegrated and the Li problem is partially solved:  ($\tau_\nuh$, $\zeta_{\nuh\rightarrow e}$) =($10^4-10^5$ s, $10^{-6}-10^{-7}$ GeV).
We also found that the sterile neutrino mass is required to be $M_\nuh \gtrsim 14$ MeV.  A lighter neutrino can not 
destroy any significant fraction of $^7$Be via photodisintegration without violating the constraints on the D abundance or the effective neutrino number.
The best parameter region is narrow even in the case of $M_\nuh \gtrsim 14$ MeV.  In this parameter region, the $^7$Be destruction by more than a factor of three can not be realized since the constraint on the D abundance excludes this possibility (Sec. \ref{sec6b}).

Third, it was found that in the best parameter region, the $\nuh$ decay 
not only decreases the $\eta$ value slightly but also increases the $N_{\rm eff}$ value by a factor of $\Delta N_{\rm eff} \lesssim$ 1.
For the moment, the 2 $\sigma$ ranges of the D abundance from QSO observations and the $N_{\rm eff}$ value from CMB observations do not indicate any effect by the sterile neutrino decay as considered in this paper.  The $\eta$ value at the cosmological recombination is consistent with the value at the BBN epoch inferred from measurements of primordial light element abundances, and the effective neutrino number is consistent with the case of only three active neutrino, i.e., $N_{\rm eff}=3$. 
Since error bars on the $\eta$ and $N_{\rm eff}$ values are getting smaller,
this model for the Li reduction can be tested by future observations of the parameters $\eta$ and $N_{\rm eff}$ (Sec. \ref{sec6b}).

Fourth, we compared results of the $\nuh$ decay with the new and old cross sections of $^7$Be($\gamma$, $\alpha$)$^3$He and $^7$Li($\gamma$, $\alpha$)$^3$H.  The new rates for the former and latter reactions are 2.3 and 2.5 times, respectively, larger than the corresponding old rates.  The corrected cross sections thus resulted in significantly smaller efficiencies of $^7$Be and $^7$Li photodisintegration.  Therefore, one should adopt the precise cross sections in calculations of nonthermal nucleosynthesis (Sec. \ref{sec6c}).

Fifth, the thermal freeze-out abundance of the sterile neutrino was estimated and compared with the best parameter region for the $^7$Li reduction.  The freeze-out abundance is much larger than the value required for the $^7$Li reduction.  Therefore, the relic sterile neutrino must be diluted before the BBN epoch by some mechanism.  A sufficiently large dilution is, however, not realized by a decay of another sterile neutrino with a mass smaller than the EW scale of $\sim 100$ GeV.  For example, therefore, other particles decaying before the BBN epoch are required for a successful $^7$Be destruction associated with the $\nuh$ decay studied in this paper (Sec. \ref{sec7}).

\acknowledgments
We are grateful to Shintaro Eijima and Kazuhiro Takeda for helpful comments.
This work was supported by Inoue Foundation for Science [HI].

\appendix
\section{formulae of sterile neutrino decay}\label{app1}
We derive the total decay rate of sterile neutrinos, and energy spectra and average energies 
of electron and positron generated in the decay.  
The overall amplitude of the matrix element squared for the decay of $\nuh_I \rightarrow \nu_\alpha + e^- + e^+$ is given by
\begin{equation}
\left|{\cal M}\right|^2=32 G_{\rm F}^2 \Theta^2 \left[
A \left(p_1 \cdot p_3 \right) \left(p_2 \cdot p_4 \right)
+B \left(p_1 \cdot p_4 \right) \left(p_2 \cdot p_3 \right)
+C m_e^2 \left(p_1 \cdot p_2 \right)  \right]
\label{eq_a1}
\end{equation}
where
$G_{\rm F}$ is the Fermi constant,
$\Theta \ll 1$ is the mixing angle,
$p_i$ is the four momentum of particle $i$,
the subscript $i$ identifies the particle species as $i=1$ for $\nuh_I$, 2 for $\nu_\alpha$, 3 for $e^-$, and 4 for $e^+$, and
constant parameters $A$, $B$ and $C$ are defined as
\begin{eqnarray}
A & = & \left(c_V + c_A \right)^2, \label{eq_a2}\\
B & = & \left(c_V - c_A \right)^2 +4 \delta_{e \alpha} +4 \left(c_V + c_A\right)\delta_{e \alpha}, \label{eq_a3}\\
C & = & \left(c_V^2 - c_A^2 \right) +2\left(c_V - c_A\right) \delta_{e \alpha},\label{eq_a4}
\end{eqnarray}
where
$c_V=-1/2+2\sin^2 \theta_{\rm W}$ and $c_A=-1/2$ are the constants for vector and axial couplings 
of charged leptons to the $Z^0$ weak boson with $\sin^2 \theta_{\rm W}=0.23$ \cite{Beringer:1900zz} the weak angle.  
The $A$ term and the first terms of $B$ and $C$ correspond to the $Z^0$ exchange, 
while the second term of $B$ corresponds to the $W^\pm$ exchange.  
The third term of $B$ and the second term of $C$ correspond to the interference contribution. When $\alpha=e$ is satisfied, parameter values are $A=0.2916$, $B=2.052$, and $C=0.6716$.

The differential decay rate as a function of energies of $e^-$ and $e^+$, i.e., $E_3$ and $E_4$, is then given by
\begin{equation}
\frac{ d^2 \Gamma}{d x_3 dx_4} = \frac{G_{\rm F}^2 \Theta^2 M_\nuh^5}{64 \pi^3} 
\left[
A x_3 \left(1 - x_3 \right) + B x_4 \left(1- x_4 \right) + 2 C x_{\rm m}^2 \left(2 - x_3 -x_4 \right) 
\right],
\label{eq_a5}
\end{equation}
where
new dimensionless variables were defined as follows: $x_{\rm m}=m_e/M_\nuh$ and $x_i=2E_i/M_\nuh$ \footnote{We note that coefficients of Eq. (A9) in Ref. \cite{Johnson1997} are erroneous.}.

The differential decay rates as a function of $x_3$ and $x_4$ are given \cite{Johnson1997} 
\footnote{Equation (A10) of Ref. \cite{Johnson1997} should be multiplied by 1/4 after coefficients are corrected.} by
\begin{eqnarray}
\frac{ d \Gamma}{d x_3} &=& \frac{G_{\rm F}^2 \Theta^2 M_\nuh^5}{64 \pi^3} 
\left\{
A x_3 \left(1 - x_3 \right) x_{{\rm f}3} + B \left(\frac{x_{{\rm f}3}^2}{2}- \frac{x_{{\rm f}3}^3}{3} \right)
+ 2 C x_{\rm m}^2 \left[\left(2 - x_3\right) x_{{\rm f}3} -\frac{x_{{\rm f}3}^2}{2} \right]\right\}_{x_{{\rm f}3}=x_{{\rm f}3,-}}^{x_{{\rm f}3,+}}, \label{eq_a6}
\\
\frac{ d \Gamma}{dx_4} &=& \frac{G_{\rm F}^2 \Theta^2 M_\nuh^5}{64 \pi^3} 
\left\{
A \left(\frac{x_{{\rm f}4}^2}{2}- \frac{x_{{\rm f}4}^3}{3} \right)
+ B x_4 \left(1 - x_4 \right) x_{{\rm f}4}
+ 2 C x_{\rm m}^2 \left[\left(2 - x_4\right) x_{{\rm f}4} -\frac{x_{{\rm f}4}^2}{2} \right]\right\}_{x_{{\rm f}4}=x_{{\rm f}4,-}}^{x_{{\rm f}4,+}}, \label{eq_a7}
\end{eqnarray}
where
$x_{{\rm f} i}$ is the variable integrated in deriving Eqs. (\ref{eq_a6}) and (\ref{eq_a7}) from Eq. (\ref{eq_a5}) ($x_{{\rm f}3}=x_4$ and $x_{{\rm f}4}=x_3$),
$x_{{\rm f}i,-}$ and $x_{{\rm f}i,+}$ are its minimum and maximum values, respectively, and
the terms in braces are evaluated as differences between values for $x_{{\rm f}i,\pm}$, i.e., $\{F(x_{{\rm f}i})\}_{x_{{\rm f}i}=x_{{\rm f}i,-}}^{x_{{\rm f}i,+}} = F(x_{{\rm f}i,+}) -F(x_{{\rm f}i,-})$.
These rates are derived by integration of Eq. (\ref{eq_a5}) over $x_{{\rm f}i}$ 
in the range of $x_{{\rm f}i,-} \leq x_{{\rm f} i} \leq x_{{\rm f}i,+}$.  On the other hand, the ranges of $x_3$ and $x_4$ 
in Eqs. (\ref{eq_a6}) and (\ref{eq_a7}), respectively, are $2x_{\rm m} \leq x_i \leq 1$.  
The values $x_{{\rm f}i,\pm}$ are given \cite{Johnson1997} by
\begin{equation}
x_{{\rm f}i, \pm} = \frac{\left(2 -x_i\right) \left( 1+ 2x_{\rm m}^2 - x_i \right) \pm \left(1-x_i \right) \sqrt[]{\mathstrut x_i^2 - 4x_{\rm m}^2}
}
{2 \left(1 +x_{\rm m}^2 -x_i\right)}.
\label{eq_a8}
\end{equation}

We define dimensionless spectra as
\begin{equation}
\frac{ d \Gamma'}{d x_i} =\left( \frac{G_{\rm F}^2 \Theta^2 M_\nuh^5}{64 \pi^3}\right)^{-1} \frac{ d \Gamma}{d x_i}. \label{eq_a22}
\end{equation}
Then, the following expressions are found,
\begin{eqnarray}
\frac{ d \Gamma'}{d x_3} &=& A f_1(x_3) + B f_2(x_3) + C f_3(x_3), 
\label{eq_a23}
\\
\frac{ d \Gamma'}{dx_4} &=& A f_2(x_4) + B f_1(x_4) + C f_3(x_4), 
\label{eq_a24}
\end{eqnarray}
where
\begin{eqnarray}
f_1(x_i) &=& x_i \left(1 - x_i \right) \left( x_{{\rm f}i,+} - x_{{\rm f}i,-}\right) \label{eq_a25}
\\
f_2(x_i) &=& \frac{x_{{\rm f}i,+}^2 - x_{{\rm f}i,-}^2}{2}- \frac{x_{{\rm f}i,+}^3 - x_{{\rm f}i,-}^3 }{3} \label{eq_a26}
\\
f_3(x_i) &=& 2 x_{\rm m}^2 \left[\left(2 - x_i\right) \left( x_{{\rm f}i,+} - x_{{\rm f}i,-}\right) -\frac{x_{{\rm f}i,+}^2 - x_{{\rm f}i,-}^2}{2} \right]. \label{eq_a27}
\end{eqnarray}

The total decay rate is given \cite{Johnson1997,Gorbunov2007} by 
\begin{eqnarray}
\Gamma (\nuh\rightarrow \nu_\alpha e^+ e^-)&=& \frac{G_{\rm F}^2 \Theta^2 M_\nuh^5}{192 \pi^3} 
\left\{
C_1 \left[ \left( 1-14 x_{\rm m}^2 -2 x_{\rm m}^4 -12 x_{\rm m}^6 \right) \sqrt[]{\mathstrut 1- 4x_{\rm m}^2} 
-12 x_{\rm m}^4 \left( 1- x_{\rm m}^4 \right) L \right] \right. \nonumber\\
&& \left. ~~~~~~~~~~~~~~~
+ 4 C_2 \left[ x_{\rm m}^2 \left( 2 + 10 x_{\rm m}^2 -12 x_{\rm m}^4 \right) \sqrt[]{\mathstrut 1- 4x_{\rm m}^2} 
+6 x_{\rm m}^4 \left( 1- 2x_{\rm m}^2 + 2x_{\rm m}^4 \right) L \right]
\right\},
\label{eq_a9}
\end{eqnarray}
where
\begin{eqnarray}
C_1 & = & \frac{A+B}{4}, \label{eq_a10}\\
C_2 & = & \frac{C}{4}, \label{eq_a11}\\
L & = & \ln \left[ \frac{1- 3x_{\rm m}^2 - \left( 1- x_{\rm m}^2 \right) \sqrt[]{\mathstrut 1
- 4x_{\rm m}^2}}{x_{\rm m}^2 \left( 1+ \sqrt[]{\mathstrut 1- 4x_{\rm m}^2} \right)} \right].
\label{eq_a12}
\end{eqnarray}
 The adopted weak angle of $\sin^2 \theta_{\rm W}=0.23$ \cite{Beringer:1900zz} corresponds to the values of $C_1=0.5858$ 
and $C_2=0.1679$ for the neutrino flavor of the final state $\alpha=e$.

The number spectra of the electron and positron emitted at the decay are given by
\begin{eqnarray}
P_{e^-}(x_3) &=& \frac{1}{\Gamma} \frac{ d \Gamma}{d x_3}, \label{eq_a13}\\
P_{e^+}(x_4) &=& \frac{1}{\Gamma} \frac{ d \Gamma}{d x_4}. \label{eq_a14}
\end{eqnarray}
The total spectra of electron and positron is given by
\begin{eqnarray}
P_{e}(x) & = & P_{e^-}(x) + P_{e^+}(x) \nonumber\\
&=& \frac{1}{\Gamma} \frac{G_{\rm F}^2 \Theta^2 M_\nuh^5}{64 \pi^3} 
\left\{
\left( A+B \right) \left[ \frac{x_{{\rm f}}^2}{2}- \frac{x_{{\rm f}}^3}{3} + x_{{\rm f}} x \left(1 - x \right) \right] 
+ 2 C x_{\rm m}^2 x_{{\rm f}} \left(4-2x-x_{{\rm f}} \right)
\right\}_{x_{{\rm f}}=x_{{\rm f},-}}^{x_{{\rm f},+}}. \label{eq_a15}
\end{eqnarray}

The average energies of electron ($E_3$) and positron ($E_4$) are given by
\begin{eqnarray}
\bar{x_3} &=& \frac{1}{\Gamma} \int_{2x_{\rm m}}^1 x_3 \frac{ d \Gamma}{d x_3} d x_3 \nonumber \\
&=& \frac{1}{\Gamma} \frac{G_{\rm F}^2 \Theta^2 M_\nuh^5}{64 \pi^3} 
f_E(A,B,C,x_{\rm m}), \label{eq_a16}\\
\bar{x_4} &=& \frac{1}{\Gamma} \int_{2x_{\rm m}}^1 x_4 \frac{ d \Gamma}{d x_4} d x_4 \nonumber \\
&=& \frac{1}{\Gamma} \frac{G_{\rm F}^2 \Theta^2 M_\nuh^5}{64 \pi^3} 
f_E(B,A,C,x_{\rm m}), \label{eq_a17}
\end{eqnarray}
where
we defined a function:
\begin{eqnarray}
f_E(A,B,C,x_{\rm m})&=& A \left\{
\frac{1}{60} \sqrt[]{\mathstrut 1- 4x_{\rm m}^2}  \left(3-29 x_{\rm m}^2 +48x_{\rm m}^4 -70 x_{\rm m}^6 -60 x_{\rm m}^8 \right) \right.
\nonumber\\
&& \left. ~~~~~
- x_{\rm m}^4 \left[ \left(1+x_{\rm m}^2\right) \left( 1-x_{\rm m}^4 \right) L_1
+\left( 3- x_{\rm m}^2 +x_{\rm m}^4 +x_{\rm m}^6 \right) L_2 \right] \right\} 
\nonumber\\
&&-B \left\{
\frac{1}{60} \sqrt[]{\mathstrut 1- 4x_{\rm m}^2}  \left(9-52 x_{\rm m}^2 +14x_{\rm m}^4 +80 x_{\rm m}^6 +120 x_{\rm m}^8 \right) \right.
\nonumber \\
&& \left. ~~~~~~~
+ x_{\rm m}^4 \left[ \left(1-x_{\rm m}^4 -2x_{\rm m}^6 \right) L_1
+\left( 3 +x_{\rm m}^4 +2x_{\rm m}^6 \right) L_2 \right] \right\}
\nonumber \\
&&+C \left\{
\frac{1}{6} x_{\rm m}^2 \sqrt[]{\mathstrut 1- 4x_{\rm m}^2}  \left(5-12 x_{\rm m}^2 +10x_{\rm m}^4 -12 x_{\rm m}^6 \right) \right.
\nonumber \\
&& \left. ~~~~~~~
+ 2x_{\rm m}^4 \left[ \left(1-x_{\rm m}^2\right) \left( 1-x_{\rm m}^4 \right) L_1
+\left( 1+ x_{\rm m}^2 +x_{\rm m}^4 -x_{\rm m}^6 \right) L_2 \right] \right\}
\nonumber \\
&&+(B - 2 C x_{\rm m}^2) \left\{
\frac{1}{24} \sqrt[]{\mathstrut 1- 4x_{\rm m}^2}  \left(5 -38x_{\rm m}^2 +6x_{\rm m}^4 +36 x_{\rm m}^6 \right) \right.
\nonumber \\
&& \left. ~~~~~~~~~~~~~~~~~~~~
- \frac{1}{2} x_{\rm m}^4 \left(1+3x_{\rm m}^4 \right) \left(L_1 -L_2 \right) \right\},
\label{eq_a18}
\end{eqnarray}
where
parameters $L_1$ and $L_2$ are defined as
\begin{eqnarray}
L_1 & = & \ln \left[ \frac{1- 3x_{\rm m}^2 - \left( 1- x_{\rm m}^2 \right) \sqrt[]{\mathstrut 1- 4x_{\rm m}^2}}{2x_{\rm m}^3} \right], \label{eq_a19}\\
L_2 & = & \ln \left[ \frac{1+ \sqrt[]{\mathstrut 1- 4x_{\rm m}^2}}{2x_{\rm m}} \right],
\label{eq_a20}
\end{eqnarray}
and $L_1-L_2=L$ is satisfied.

\begin{figure}
\begin{center}
\includegraphics[width=8.0cm,clip]{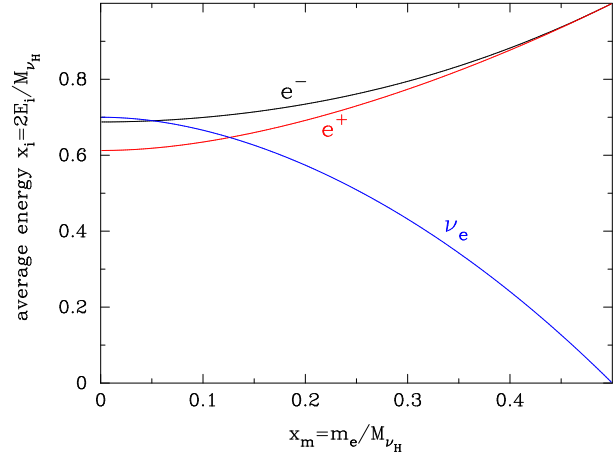}
\caption{Average energies of electron, positron and neutrino generated at the $\nuh$ 
decay as a function of the mass ratio of electron and sterile neutrino. \label{fig7}}
\end{center}
\end{figure}

Figure \ref{fig7} shows average energies of electron, positron and neutrino generated at the $\nuh$ 
decay as a function of $x_{\rm m}$ calculated with Eqs. (\ref{eq_a16}) and (\ref{eq_a17}) and 
a trivial relation of $\bar{x_2}=2-(\bar{x_3}+\bar{x_4})$.  In the small $x_{\rm m}$ region, 
average energies of all the three particles in the final state are close to one third 
of the sterile neutrino mass $M_\nuh$, i.e., $\bar{x_i} \sim 2/3$.  In the large $x_{\rm m} (\lesssim 1/2)$ region, 
on the other hand, masses of electron and positron are significant fractions of the sterile neutrino mass.  
The most of the energy in the final state is, therefore, taken for the mass energy, and the average energy of $\nu_e$ is small.


Finally, the decay rate for the mode of $\nuh_I \rightarrow \sum_\beta \nu_e \bar{\nu_\beta} \nu_\beta$ is given \cite{Gorbunov2007} by
\begin{eqnarray}
\Gamma (\nuh\rightarrow \sum_\beta \nu_e \bar{\nu_\beta} \nu_\beta)&=& \frac{G_{\rm F}^2 \Theta^2 M_\nuh^5}{192 \pi^3}.
\label{eq_a21}
\end{eqnarray}
We note that the decay into the final state of $\nu_\alpha + \bar{\nu_\beta} + \nu_\beta$ for 
$\alpha=\mu$ and $\tau$ does not occur in the assumption adopted in this paper (see Sec. \ref{sec2c}).



\end{document}